\documentclass[11pt,preprint,graphicx]{aastex}
\usepackage{graphicx}

\def\etal{\it et al. \rm }

\begin{document} 

\title{Stellar Populations and the Star Formation Histories of LSB Galaxies:
V. WFC3 Color-Magnitude Diagrams}

\author{James Schombert}
\affil{Department of Physics, University of Oregon, Eugene, OR 97403;
jschombe@uoregon.edu}

\author{Stacy McGaugh}
\affil{Department of Astronomy, Case Western Reserve University, Cleveland, OH 44106;
stacy.mcgaugh@case.edu}

\begin{abstract}

\noindent We present WFC3 observations of three low surface brightness (LSB) galaxies
from the Schombert \etal LSB catalog that are within 11 Mpc of the Milky Way.  Deep
imaging at F336W, F555W and F814W allow the construction of the $V-I$
color-magnitude diagrams (CMD) to $M_I = -2$.  Overall 1869, 465 and 501 stellar
sources are identified in the three LSB galaxies F415-3, F608-1 and F750-V1.  The
spatial distribution of young blue stars matches the H$\alpha$ maps from ground-based
imaging, indicating that star formation in LSB galaxies follows the same style as in
other irregular galaxies.  Several star complexes are identified, matching regions of
higher surface brightness as seen from ground-based imaging.  The color-magnitude
diagrams for each LSB galaxy has the similar morphology to Local Volume (LV) dwarf
galaxies, i.e. a blue main sequence, blue and red He burning branches and asymptotic
giant branch (AGB) stars.  The LSB CMD's distinguish themselves from nearby dwarf
CMD's by having a higher proportional of blue main sequence stars and fewer AGB stars
than expected from their mean metallicities.  Current [Fe/H] values below $-$0.6 are
deduced from the position of the red helium-burning branch (rHeB) stars in the $V-I$
diagram.  The distribution of stars on the blue helium-burning branch (bHeB) and rHeB
from the $U-V$ and $V-I$ CMD indicate a history of constant star formation for the
last 100 Myrs.

\end{abstract}

\section{Introduction}

The importance of low surface brightness (LSB) galaxies to galaxy studies is not that
they dominate the total galaxy population of the Universe (they do not, Hayward,
Irwin \& Bregman 2005; Rosenbaum \& Bomans 2004; Schombert, Pildis \& Eder 1997) nor
that they represent a special form of star formation (they do not, Schombert \&
McGaugh 2014).  Instead, their importance lies in the need to explore the full range
of galaxy characteristics in order to derive formation and evolutionary scenarios
unbiased by size, mass or density.  A clear picture of galaxy formation requires the
inclusion of the LSB realm both globally, those galaxies that are LSB in their mean
stellar densities (Pildis, Schombert \& Eder 1997), and locally, the LSB regions of a
galaxy (Boissier \etal 2008)

This series of papers (Schombert, Maciel \& McGaugh 2011, Schombert, McGaugh \&
Maciel 2013, Schombert \& McGaugh 2014a, Schombert \& McGaugh 2014b) has explored the
class of LSB galaxies selected from visual survey only by their mean surface
brightness (in this sense, a class of objects that occupy the faintest end of central
surface brightness distribution).  The LSB class contains a full range of galaxy
types (irregulars to disks) and a full range of size and luminosity (dwarfs to
giants).  They are found in all types of galaxy environments (Schombert \etal 1992;
van Dokkum \etal 2014), but tend to avoid the dense, rich environments, such as
cluster cores (Galaz \etal 2011).  While LSB dwarf ellipticals (dE's) and dSph are
gas-poor and often found in clusters, the typical LSB galaxy is gas-rich (Huang \etal
2014, and references therein), but with low H$\alpha$ fluxes indicating low current
star formation rates (SFR; Schombert, Maciel \& McGaugh 2011).

Over the years, and numerous studies by many observing teams, the following 
characteristics were found to be in common with galaxies at the low end of
the surface brightness spectrum.  There are, of course, exceptions to all of the
following generalizations, but a majority of LSB galaxies maintain these trends.
First, the ratio of gas to stellar masses increases dramatically with decreasing mean
surface brightness such that the highest gas fraction galaxies ($M_{gas}/(M_{gas}+M_*)$)
are typically LSB as has been shown in previous studies (McGaugh \& de Blok
1997, Geha \etal 2006).  This is not too surprising in that baryonic content of galaxies
is primarily stars and gas, so the decreased importance of stellar mass will
naturally result in an increasing fractional dominance of gas.

Second, the optical and near-infrared (near-IR) colors of LSB galaxies are atypically blue (Pildis,
Schombert \& Eder 1997; Schombert \& McGaugh 2014b).  This is only considered atypical
in the sense that their current SFR are extremely low, with values in common with red
spirals and S0's.  However, galaxy colors have a trend of becoming bluer with more
irregular morphology and higher gas fractions, so the expectation, just based on
morphology and gas content, is that LSB galaxies should have similar colors as
other irregular galaxies.

Third, unsurprisingly, LSB galaxies do have extremely low current SFRs (on average a
factor of ten lower than other irregular galaxies of similar stellar mass; Schombert,
McGaugh \& Maciel 2013).  The style of star formation is similar to other irregular
galaxies, i.e., most of the star formation is concentrated in HII regions.  This
dispels a scenario where LSB galaxies only form stars outside molecular complexes
(Schombert \etal 1990).  LSB galaxies also have a normal range of HII region sizes
from massive $10^5$ $M_{\sun}$ complexes to individual OB star regions (Schombert,
McGaugh \& Maciel 2013), i.e., massive complexes do occur despite their low mean
stellar densities.

The key dilemma from LSB galaxies studies is primarily concerning their stellar
populations.  As the mean surface brightness decreases, the optical and near-IR
colors become bluer, SFR decreases.  This is opposite to the expectation that a galaxy
whose current SFR's are low should be dominated by the older, redder population,
particularly in the lowest surface brightness regions.  Contrary, Schombert, McGaugh
\& Marciel (2013) found that the optically dimmest regions were, in fact, the bluest
regions.  Near-IR imaging revealed that the colors for LSB galaxies could be
explained by a combination of metallicity and the pattern of recent star formation
(low level bursts separated by quiescent epochs; Schombert \& McGaugh 2014a).  Models
of roughly constant mean star formation rate punctuated by stochastic variations in
current SFR also agree well with constraints from kinematic studies, where LSB
irregulars display solid body rotation rather than the differential rotation that
drives most spiral patterns (Kuzio de Naray, McGaugh \& Mihos 2009, Eder \& Schombert
2000).  Constant star formation, in turn, provides a natural explanation for the
observed range of stellar mass-to-light ratios in LSB galaxies (Schombert \& McGaugh
2014a).

A more direct method to understanding the star formation history of LSB galaxies is
to, of course, resolve the stellar population using deep space imaging (Dalcanton
\etal 2009).  For even the top of a stellar population's color-magnitude diagram
(CMD) reveals detailed information on the evolution of the underlying stars, the SFR
over the last Gyr and chemical evolution of those stars.  Interpretation is assisted
by numerous synthetic CMD simulators that use the most recent stellar isochrones and
take into account short-lived, but highly luminous, phases of stellar evolution (e.g.,
asymptotic giant branch stars, blue stragglers, etc., see Gallart \etal 1996; Mighell
1997; Holtzman \etal 1999; Dolphin 2002; Aparicio \& Hidalgo 2009)

The analysis of CMD's in numerous Local Volume (LV, galaxies within 10 Mpc
of the Milky Way) primary and dwarf galaxies has
filled the literature in the last ten years (see Tully \etal 2009; McQuinn \etal
2012).  These studies have found that the star formation history of most LV dwarfs is
a complicated progression of bursts of star formation; however, they all also have
significant old stellar populations ($\tau >$ 8 Gyrs) that dominate their global
colors and the lower portion of the CMD.  Resolved populations also give spatial information
on the star formation history (SFH) of LV dwarfs (McQuinn \etal 2012), usually solid
body rotation prevents significant mixing on 50 Myr timescales (Bastian \etal 2009)
allowing for the potentially mapping of the spatial chemical history of stars rather
than gas metallicity from emission lines (Gallart \etal 2005).

Of course, the less distant a galaxy is to the Milky Way, the fainter into a CMD one can
resolve.  Unfortunately, none of the LSB dwarfs from our samples are closer than 8
Mpc, the outer edge of successful HST stellar photometry and, thus, our
interpretation will be limited to the bright portion of the CMD.  For this study, we
have selected three dwarfs from our LSB catalogs that are the closest to the Milky
Way (F415-3, F608-1 and F750-V1) for WFC3 imaging in three filters (F336W, F555W and
F814W).  Our objective is to obtain the first observations of the top of the CMD in
any LSB galaxy for comparison to other LV dwarfs.  These galaxies also have
matching ground-based optical, near-IR, H$\alpha$ and HI imaging for
comparison to the resolved stellar populations.  Our goal is a first look
at the details of the star formation history of these low density galaxies
and their relevance to galaxy formation and evolution scenarios.

\section{Analysis}

\subsection{Sample}

Three LSB galaxies were selected from dwarf LSB catalog of Schombert, Pildis \& Eder
(1997), an optically selected sample with Arecibo HI observations.  The criteria for
inclusion in the Schombert dwarf LSB sample was 1) LSB nature, 2) one arcmin or
greater angular size and 3) irregular morphology.  The intent of the sample was to
extend the local dwarf galaxy sample as a test of biased galaxy formation scenarios.
There was no attempt to be luminosity or mass complete; size, morphology and LSB
appearance were the main criteria.

The three galaxies selected were all Fall objects (F415-3, F608-1 and F750-V1) and
all located less than 11 Mpc away.  Their distances are from HI observations (Eder \&
Schombert 2000), corrected to the CMB reference frame by NED, and are 10.4, 8.9 and 7.9 Mpc
respectively with an accuracy of 0.1 Mpc.  These were the closest LSB dwarfs in the
Schombert LSB catalog to maximize the detection of stellar populations and depth to
the resulting CMD.  F608-1 is also UGC159, where the original coordinates were
misprinted on POSS SAO overlays resulting in separate designations for several years.
We maintain the LSB catalog labels for clarity.

\begin{figure}[!ht]
\centering
\includegraphics[scale=0.85,angle=0]{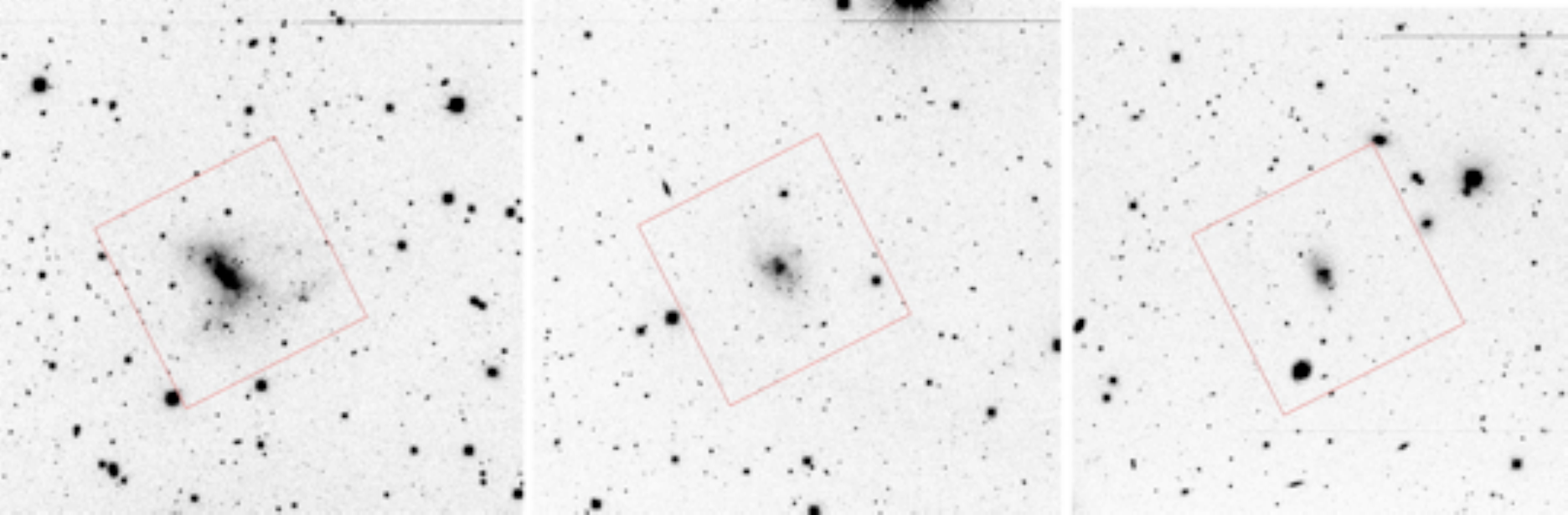}
\caption{\small The LSB galaxies F415-3 (left), F608-1 (middle) and F750-V1 (right)
from 900 sec KPNO 2.1m $V$ images.  The 3x3 arcmin WFC3 field is shown as an outline
on the ground-based images.  All three galaxies are between 8 and 11 Mpc in distance,
so each frame is approximately 20 kpc on a side.  The typical irregular morphology to
LSB dwarfs is obvious, H$\alpha$ maps of the same galaxies are found in Schombert,
McGaugh \& Maciel (2013).
}
\label{ground}
\end{figure}

Ground-based $V$ images from KPNO 2.1m for all three galaxies are shown in Figure
\ref{ground} (each image is 900 secs of exposure, 0.61 arcsecs/pixel plate scale).
Their optical and HI properties are summarized in Table 1.  Compared to other LSB
galaxies studied in H$\alpha$ (Schombert, McGaugh \& Maciel 2013), F608-1 and F750-V1
are on the small size (less than 0.6 kpc in radius at the 26 $V$ mag arcsec$^{-2}$
isophote), F415-3 is 1.8 kpc in radius, larger but still dwarf-like in size.  Their
central surface brightness are average with respect to the LSB sample as a whole,
ranging around 23 $V$ mag arcsecs$^{-2}$. F608-1 and F750-V1 have the lowest baryon
masses (gas plus stellar mass) in the LSB sample (10$^{7.8}$ and 10$^{7.3}
M_{\sun}$).  F415-3 has an average LSB mass and luminosity (10$^{8.8} M_{\sun}$).
All three have high gas fractions ($M_{gas}$/$M_*$+$M_{gas}$) greater than 70\%.

With respect to star formation, all three galaxies have very low current star
formation rates based on H$\alpha$ measurements.  They lie in the lower 10\% of the
star formation rates (SFR) for LSB galaxies (110 objects) studied by Schombert,
McGaugh \& Maciel (2013) and in the bottom 5\% of star forming dwarf galaxies (168
objects, van Zee 2001; Hunter \& Elmegreen 2004).  Despite the low SFR's, their
optical colors are very blue, typical of values for LSB galaxies and star-forming
spirals, although the blue colors in star forming galaxies is presumingly due to a
large high-mass stellar population.  The origin of blue colors in LSB galaxies is
unclear as the current SFR is low.  Possible explanations for the blue colors of LSB
has ranged from extremely low metallicities (Schombert \etal 1990) to unusual stellar
types (i.e. over abundance of bHB or blue straggler stars, Rakos \& Schombert 2004),
but no particular idea has gained support from observations.

\subsection{Observations}

The CMD's presented herein are produced by performing HST stellar photometry taken
with the Wide Field Camera 3 (WFC3), the fourth generation UV/IR imager.  WFC3 uses
two backside illuminated 2Kx4K CCDs with a combined field of view of 162x162 arcsecs.
The plate scale is 0.04 arcsecs/pixel.  Observations were taken in three filters
F336W, F555W and F814W, approximately Johnson $U$, $V$ and $I$.  These filters were
selected to isolate a CMD in $V-I$ to measure mean metallicity of the older
population and $U-V$ observations to identify the UV sources of the limited H$\alpha$
emission seen in these LSB dwarf galaxies.

Ten orbits from cycle 20 were assigned to each object, split as five orbits for F336W,
three orbits for F555W and two orbits for F814W.  Due to their low surface brightness
nature, the observations were made in LOW-SKY mode where the zodiacal light is less
than 30\% of the minimum.  Each orbit was broken into four exposures using
UVIS-DITHER-BOX for cosmic-ray subtraction and to minimize pixel-to-pixel sensitivity
variations.  A pre-flash option was used for the F336W exposures to avoid a known WFC3
charge transfer problem (Anderson \& Baggett 2014).  Orbital variations resulted in
total F336W exposure times of 8,800 secs for F415-3 and 10,000 secs for F608-1 and
F750-V1.  F555W and F814W received total exposures of 7,088 and 4,688 secs for all
three galaxies.

Reduction images were taken directly from the STScI pipeline where bias, flat-fielding
and image distortion were automatically corrected.  Calibration was performed using
standard WFC3 header values.  CMD's for all three colors, in the HST filter
system, are shown in Figure \ref{cmd}.  Conversion from the WFC3/UVIS filter
passbands to the Johnson/Cousins $UVI$ passbands was accomplished using the
photometric transformation to $AB$ magnitudes, then from $AB$ to $UVI$.  The $AB$
conversion for $V$ and $I$ are well known, but the $AB$ conversion to $U$ is less
well established.  After some investigation, we used the $AB$ and $U$ magnitude of
the Sun (6.35 and 5.61 at 336 nm, Blanton \etal 2007) for the conversion.  We
confirmed the $V$ and $I$ calibration by comparison to ground-based images calibrated
with Landolt standards.  The mean difference between HST and ground-based magnitudes
was 0.01$\pm$0.05 in $V$ and 0.02$\pm$0.07 in $I$, effectively zero.  Galactic
extinction was applied to the final photometry following the prescription of Schlafly
\& Finkbeiner (2011).  These values were 0.27, 0.15 and 0.25 for our three galaxies.

Stellar sources were identified using a threshold filter that used local sky for
discrimination.  Crowded fields in the galaxies' cores were isolated by visual
inspection; however, the crowding was by no means as intense as often appears in high
surface brightness galaxies.  Star clusters were identified by visual inspection and
assigned a letter designation.  Many of these corresponded to regions of enhanced
surface brightness in ground-based images ($V$ knots and HII regions).  Comparison
between frames made use of the internal WCS for the WFC3 frames.  Comparison to
ground-based optical and H$\alpha$ images used a coordinate system based on a dozen
stars in common with the WFC3 fields.

Stellar photometry was performed using the version of DAOPHOT found in the recent
PyRAF platform.  Although more sophisticated photometry programs were available, the
low number of sources and wide spacing indicative to LSB galaxies made for
uncomplicated PSF fitting and local sky determination.  Targets were identified by a
combination of threshold filtering and visual inspection.  A series of 10 to 15 stars
were selected as PSF standards.  The FWHM was consistent at 0.092 arcsecs in each
filter.  Non-stellar objects, based on profile sharpness, were rejected as presumed
background galaxies (although the possibility that these objects are planetary nebula
is not excluded).  A few interesting (i.e., very bright) stars were too crowded to
process automatically and were reduced by interactive tools.

Blending is a serious problem at distances of the three galaxies in our sample.
However, there are couple factors that work to minimize the effects of blending on
our results.  First, the stellar densities of our LSB galaxies are, by definition,
much lower than other LV dwarfs with resolved stellar populations.  Aside from a few
very dense regions associated with the high H$\alpha$ signature of strong star
formation, most of the stellar sources are spatially well defined.  Second, blending
by binary pairs is less of a concern than other CMD studies for we only sample the
top of the luminosity function.   Odds are that the companion star for a binary
system will be much less luminous than the primary, that have a small contribution to
the measured color.

\begin{figure}[!ht]
\centering
\includegraphics[scale=0.45,angle=0]{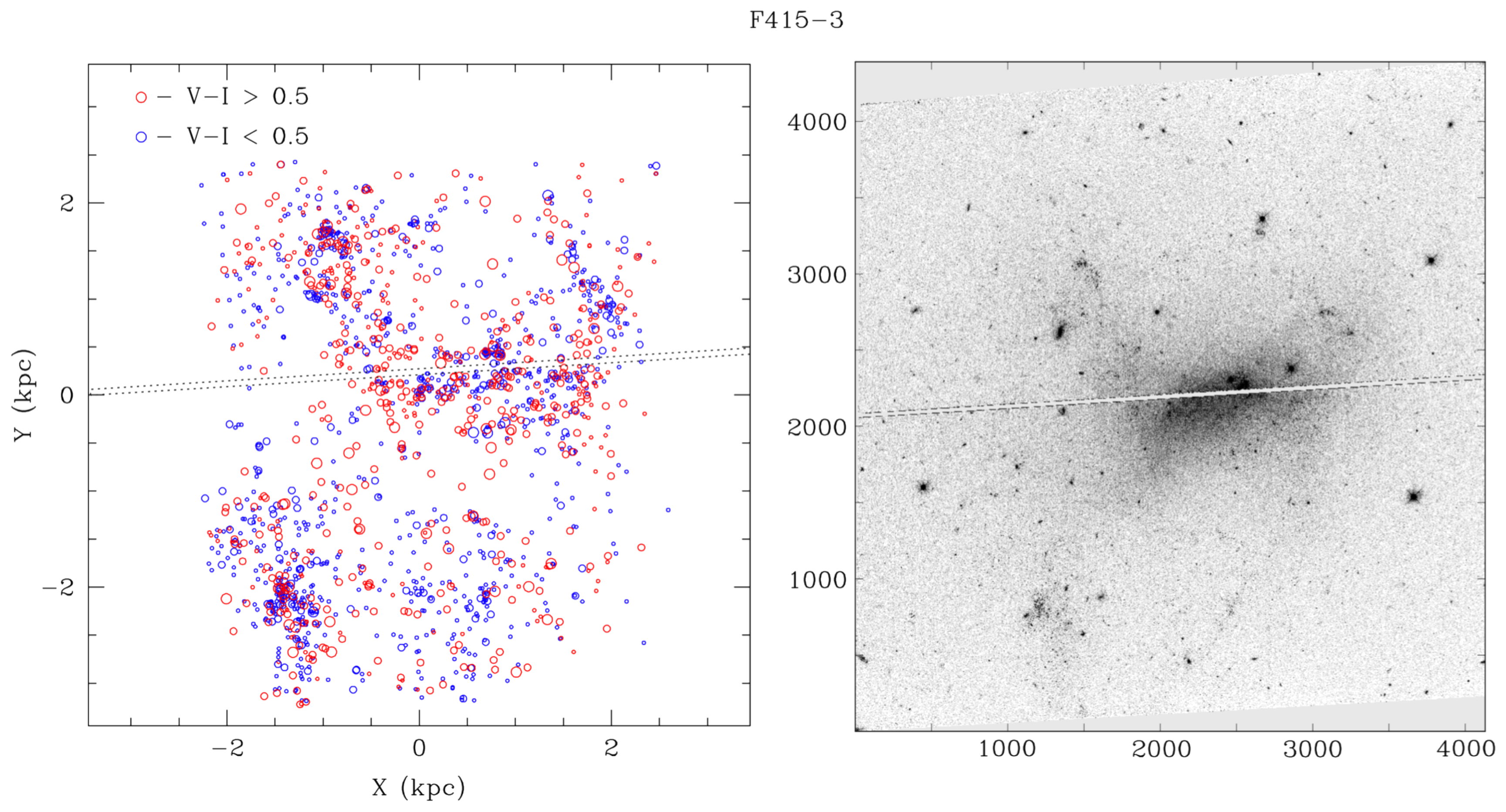}
\caption{\small The right frame displays the WFC3 F555W image (7,088 sec exposure) for
the LSB galaxy F415-3.  North is roughly towards the upper right corner, East is 90
degrees counter-clockwise.  
The left frame is a
map of 1,869 sources with S/N $>$ 5 in the F555W frame.  
Pixel units in X and Y are shown on the right, kpc's on the left.  
Symbol size correlates with luminosity, color displays blue or red based on a
$V-I=0.5$ cut.  The number counts for the stellar sources traces the ground-based
optical surface brightness.  Blue stars tend to be concentrated in clusters; however,
a significant fraction are distributed in low surface brightness regions explaining
the long-standing dilemma of the lack of sharp 2D color discrimination in LSB
galaxies.
}
\label{map1}
\end{figure}

\begin{figure}[!ht]
\centering
\includegraphics[scale=0.45,angle=0]{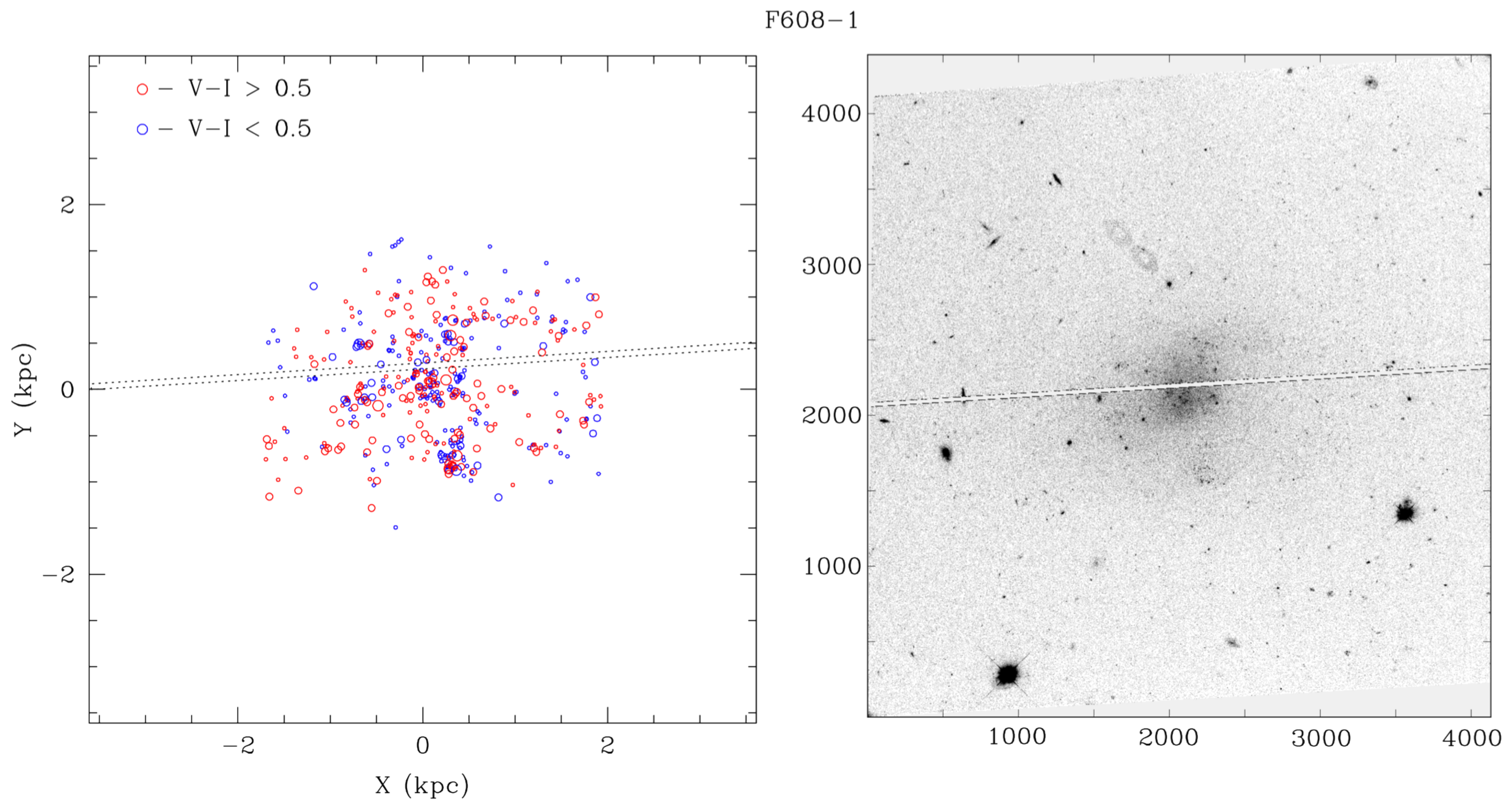}
\caption{\small The right frame displays the WFC3 F555W image (7,088 sec exposure) for
the galaxy F608-1.  North is roughly towards the upper right corner, East is 90
degrees counter-clockwise.  
The left frame is a
map of 465 sources with S/N $>$ 5 in the F555W frame.  Axes are marked in kpc's.
Pixel units in X and Y are shown on the right, kpc's on the left.  
Symbol size correlates with luminosity, color displays blue or red based on a
$V-I=0.5$ cut.  Although smaller in size that F415-3, the stellar distribution is as
extended as F415-3.  Again, the bright blue stars are associated with clusters.
}
\label{map2}
\end{figure}

\begin{figure}[!ht]
\centering
\includegraphics[scale=0.45,angle=0]{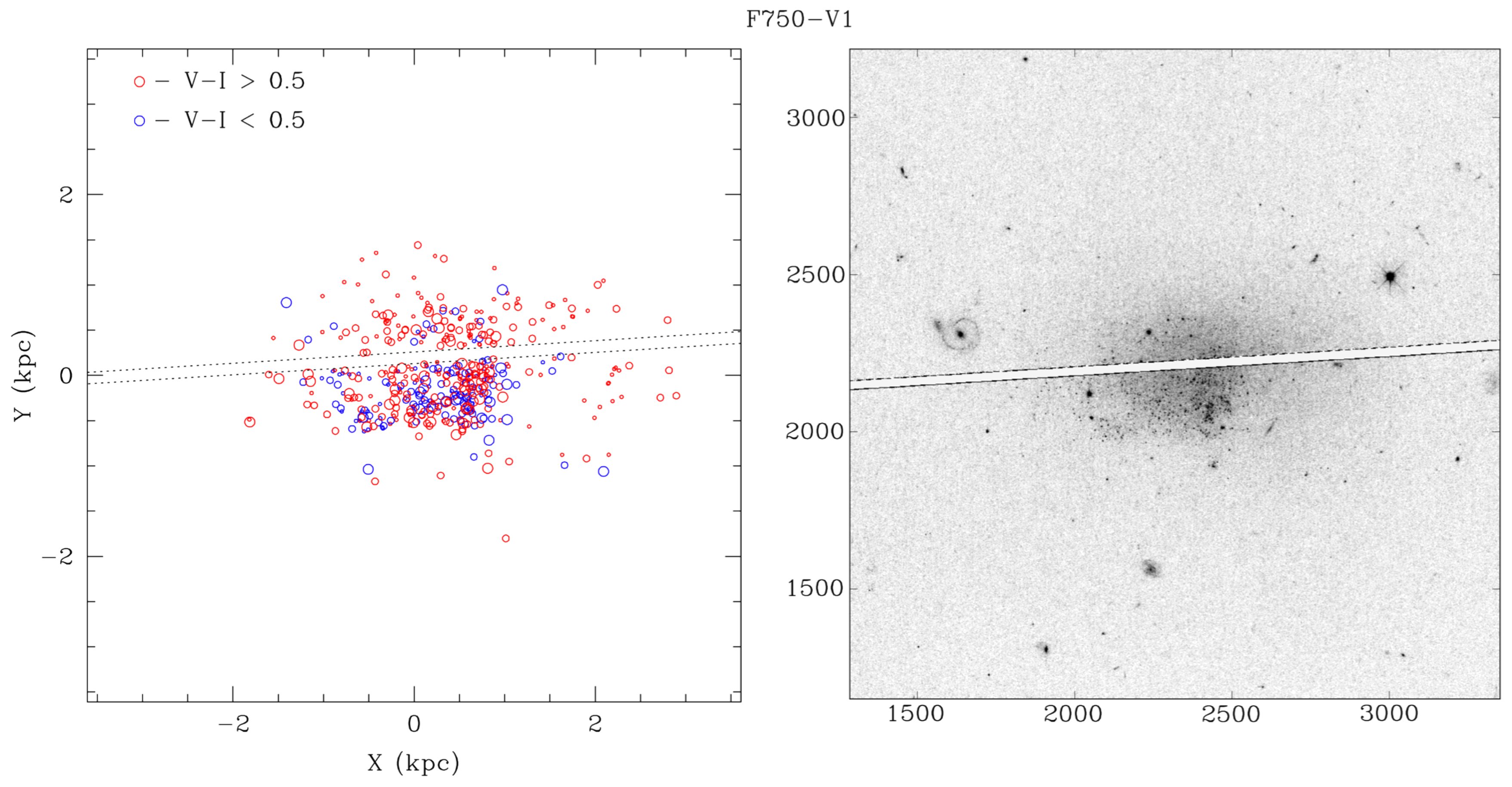}
\caption{\small The right frame displays the WFC3 F555W image (7,088 sec exposure) for
the galaxy F750-V1.  North is roughly towards the upper right corner, East is 90
degrees counter-clockwise.  The left frame is a
map of 501 sources with S/N $>$ 5 in the F555W frame.  
Pixel units in X and Y are shown on the right, kpc's on the left.  
Symbol size correlates with luminosity, color displays blue or red based on a
$V-I=0.5$ cut.  F750-V1 is smaller than F415-3 or F608-1 as the adjusted X/Y scale
indicates.
}
\label{map3}
\end{figure}

A large occulted region passes through the center of the WFC3 frame as an artifact of
the interface between the two CCD's.  This region, while manually removed from the
image, unfortunately, also passes through the center of each galaxy.  However,
offsets would have moved interesting outer star-forming regions off field.  As the
occulting strip varies in sky position from filter to filter (orbit visit to visit),
some clusters in the galaxy cores were only observed through one or two filters.  A
map of all the sources with at least two color photometry is found in Figures
\ref{map1}, \ref{map2} and \ref{map3}.

\begin{figure}[!ht]
\centering
\includegraphics[scale=0.75,angle=0]{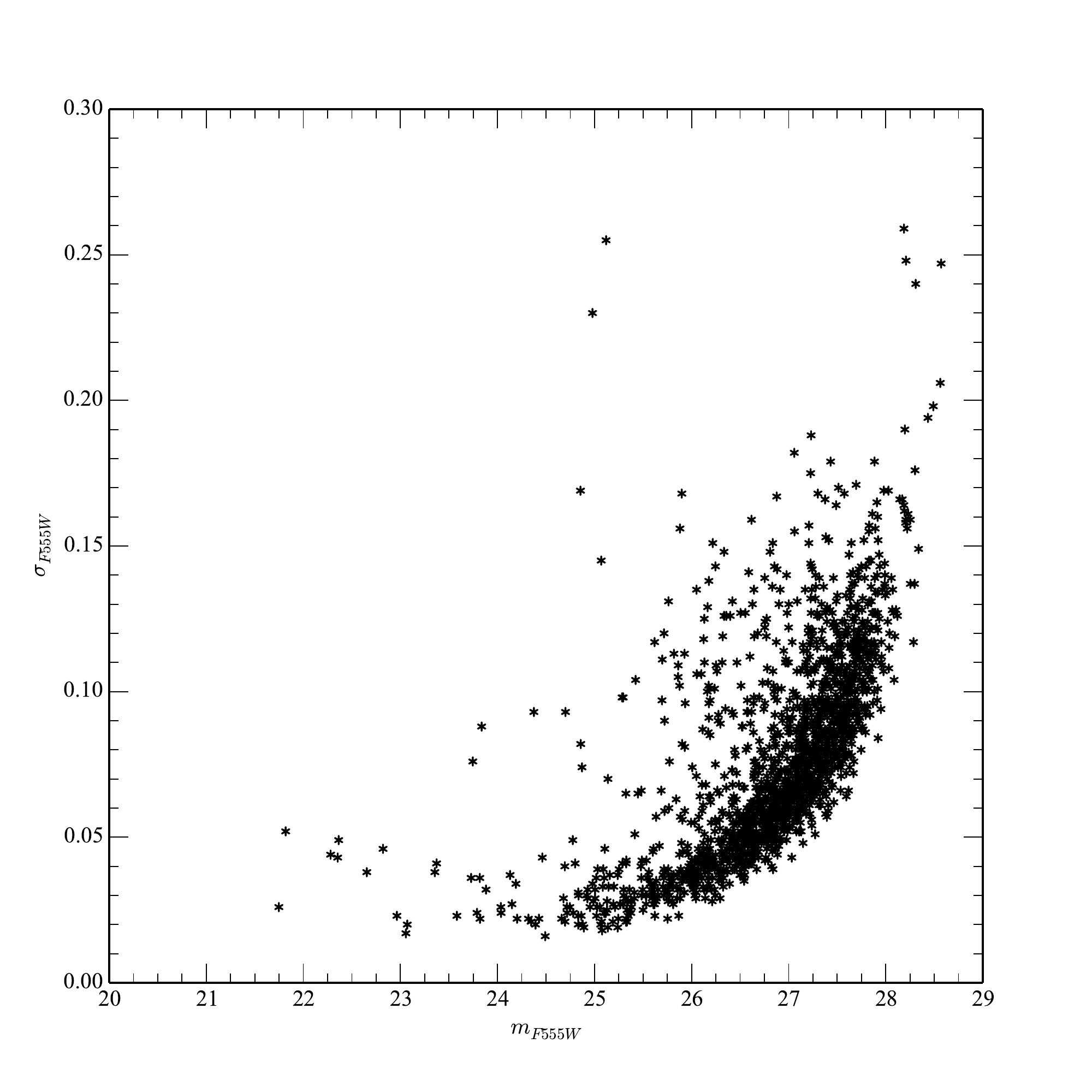}
\caption{\small Photometric errors for stellar sources in the F555W frame of F415-3.
Stars with anomalous errors are typically associated with crowded regions (poor local
sky) or asymmetric PSF shapes.  The typical error for a limiting magnitude of 26.8 is
0.07.
}
\label{limiting_mag}
\end{figure}

The limiting magnitudes and photometric errors were similar from galaxy to galaxy
since the exposure times and instrument set-up were identical.  The limiting
magnitudes were 27.1 in F336W ($U$), 27.4 in F555W ($V$) and 27.5 in F814W ($I$), which
corresponds to approximately $M_V=-2.5$ at the distances of the sample.  These were
exactly the expected limiting magnitudes based on pre-observation calculations from
HST's APT for the requested orbits and filters.  The photometric errors as a function
of F555W magnitude is shown in Figure \ref{limiting_mag}.  At the limiting magnitude,
errors reach 0.12, 0.07 and 0.20 in F336W, F555W and F814W.  Stars with errors greater
than these values were inspected visually for reality.  Experiments with artificial
stars demonstrated that our sample is complete to 80\% at the limiting magnitude in
F555W.  Thus, we use the catalog of F555W detections as the primary catalog and search
for stars detected only in F336W or F814W as extreme blue or red objects.  A total of
2,155 sources are found for all three galaxy with S/N $>$ 5 (1,869 in F415-3, 465 in
F608-1 and 501 F750-V1).  All the photometry and images can be found at our
LSB website (http://abyss.uoregon.edu/$\sim$js/lsb).

\subsection{Surface Brightness Mapping}

One of the many paradoxes for LSB galaxies is the origin of their LSB nature.  There
is the original problem of why their are so low in surface brightness as a class of
objects, compounded with the dilemma that even their lowest LSB regions are atypically
blue in color (Schombert, McGaugh \& Maciel 2013).  A faded (i.e., old) stellar
population would exhibit a LSB nature, but would be quite red (Rakos \& Schombert
2005).  Alternatively, the spacing between stars might be much larger than
star-forming spirals, resulting in less stellar luminosity per pc$^2$ or the galaxies
might be much thinner than other dwarf galaxies.  Both scenarios require very
different star formation mechanisms than the usual molecular cloud collapse into star
clusters once a critical gas density is reached (i.e., Schmidt's law, Lada 2014).

The distribution of stellar sources is the first look we have into the tip of the
underlying stellar population in LSB galaxies.  There are 1,869 detected stellar
objects in F415-3.  They range in absolute luminosity from $-$10 to $-$0.5 $I$ mag.
The total luminosity of the stellar sources is log $L/L_{\sun}$ = 6.86, compared to
log $L/L_{\sun}$ = 7.85 for the galaxy's total $V$ luminosity (using 4.83 as the
absolute magnitude of the Sun).  This means that the observed WFC3 stellar population
is only 1/10th of the total luminosity of the galaxy, the rest contributed by an
unresolved, underlying stellar population.

\begin{figure}[!ht]
\centering
\includegraphics[scale=0.8,angle=0]{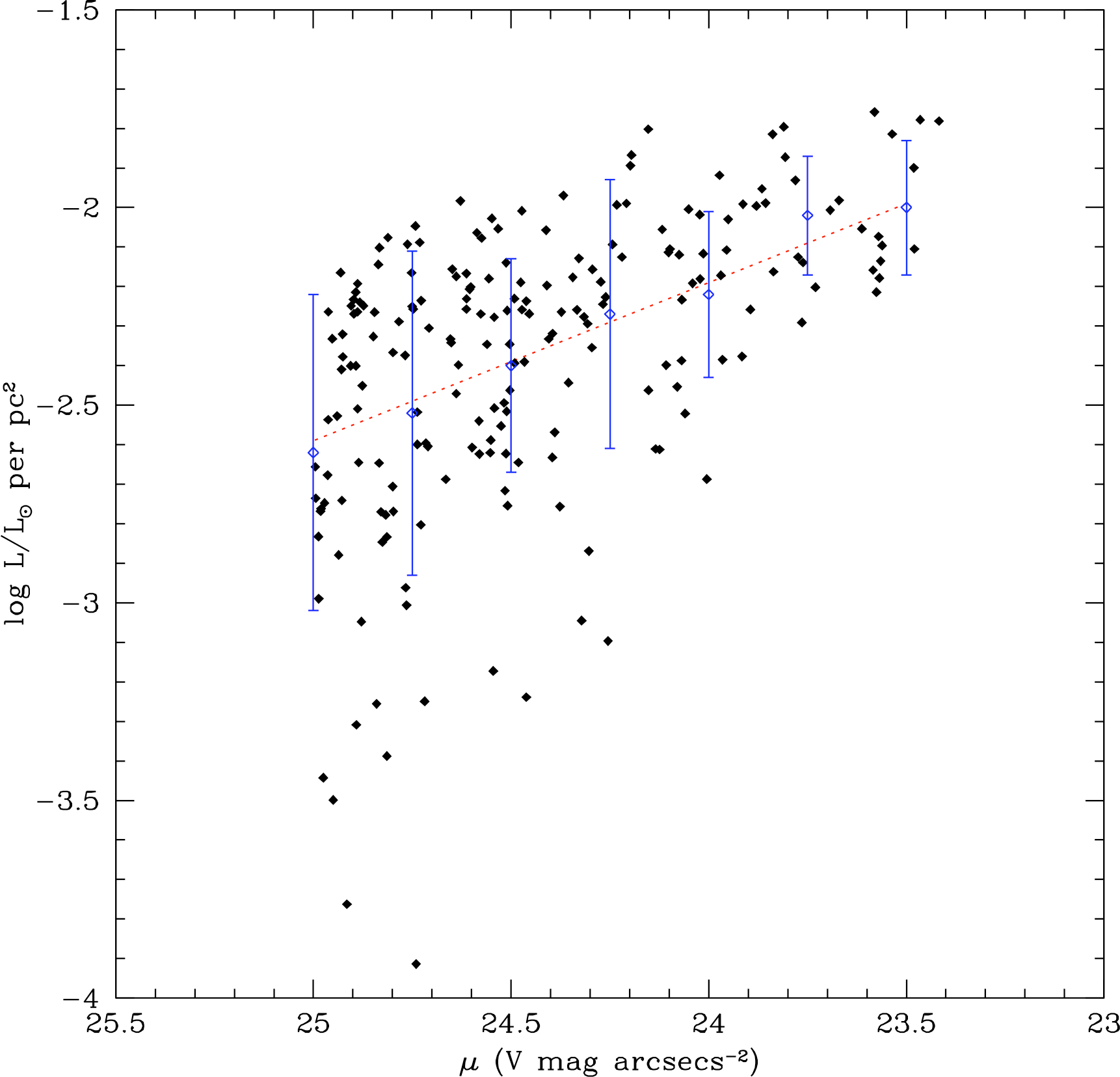}
\caption{\small Mean surface brightness versus stellar luminosity for F415-3.  The
local surface brightness in three arcsecs boxes is plotted against the total luminosity
of stellar sources in the same area.  The mean surface brightness is taken from
ground-based $V$ images, correlated against WFC3 stellar counts.  Stellar counts in
the same regions are converted to luminosities per pc$^2$.  A moving average is shown
as blue symbols.  The red dotted line is the canonical relationship between $V$
surface brightness and solar luminosities per pc$^2$ shifted by a factor of 10 to
account for the difference between the galaxy total luminosity and the sum of the
stellar counts (i.e. the unresolved stellar population).}
\label{sfb_map}
\end{figure}

We can check the distribution of surface brightness from ground-based images, which
measures the contribution from all the stars, with the luminosity distribution of the
WFC3 sources.  The mean surface brightness is taken from re-registered ground-based
images where a 3x3 arcsecs box was used to smooth the cleaned image (foreground and
background objects removed).  This is compared to the sum of the luminosity of all
the stellar sources in the same region, converted to luminosity per pc$^2$.  The
resulting correlation between the source distribution and surface brightness shown in
Figure \ref{sfb_map}, scaled by a factor of 1/10 for the luminosity of the stellar
sources.

There are several interesting points to extract from this Figure.  First, as was
expected by visual inspection of Figure \ref{map1}, there is a correlation between
stellar counts and the underlying surface brightness of the galaxy.  Regions of
bright surface brightness (knots from Paper II) are clearly associated with stellar
associations.  Regions of densely packed stellar sources are also higher in mean surface
brightness.  Note that this correlation does not necessarily have to exist, for while
the higher surface brightness regions would be associated with new star formation,
the lower surface brightness regions could be faded populations devoid of bright
stars.  This increases the confidence that conclusions based on the top of the
stellar luminosity function can be extended to the underlying stellar population.

The correlation between stellar sources and mean surface brightness also follows the
trend expected for converting surface brightness into stellar luminosity per pc$^2$
(dotted line in Figure \ref{sfb_map}, corrected for missing luminosity of undetected
stars).  A majority of the data is within the expectation of liner correlation
between surface brightness and stellar counts. This implies that for every square
parsec the relationship between the bright stars and faint, unresolved, stars is
constant.  That is, there are no hidden populations in LSB galaxies.  Aside from the
very brightest, short-lived blue stars, the other stellar sources trace the faintest,
unresolved stars as well.  The broad distribution of blue stars indicates a great
deal of uniformity by age to the stellar populations in LSB galaxies.  There appear
to be no regions that are strictly old (greater than 3 Gyrs) stars.  In most
galaxies, an old population is associated with some central concentration of light.
The irregular morphology of LSB galaxies means the older populations, if they exist,
are intermixed with the new stars.

\subsection{Star Clusters Identification}

\begin{figure}[!ht]
\centering
\includegraphics[scale=0.8,angle=0]{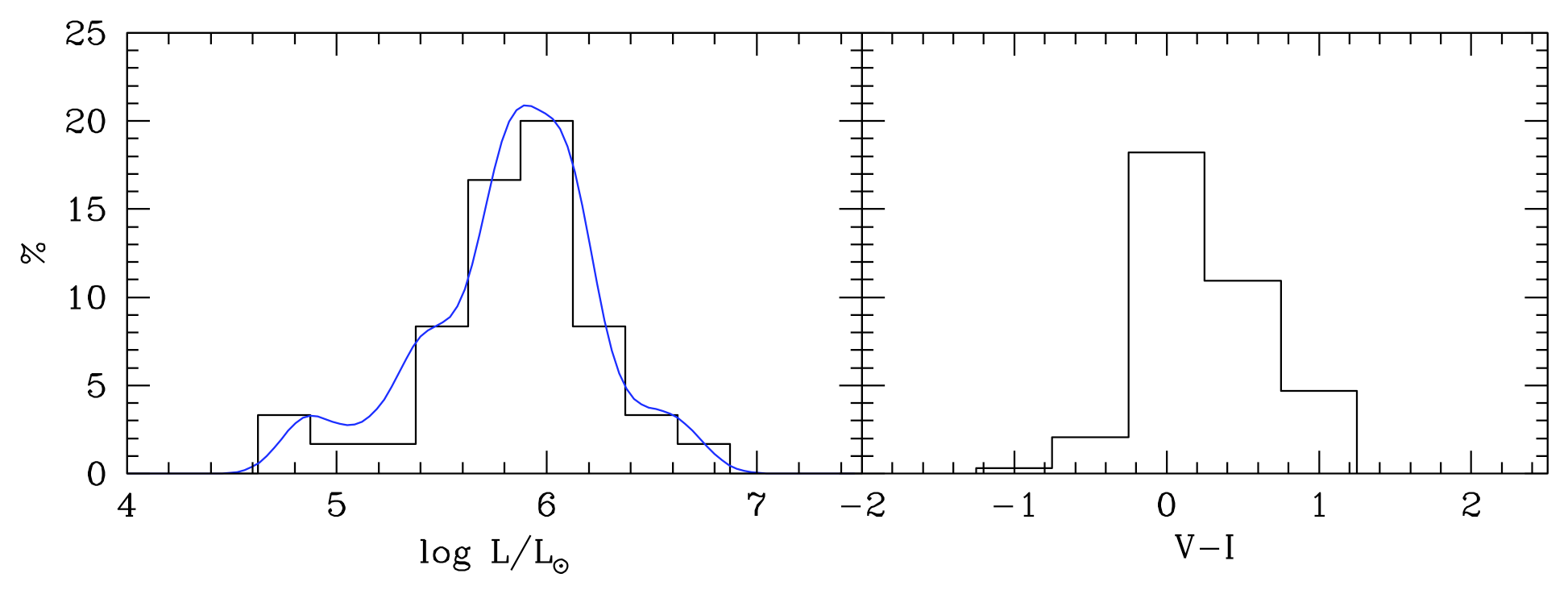}
\caption{\small Histograms of the total $V$ luminosities and weighted $V-I$ colors of the 39
star groupings identified in all three galaxies.  These values are consistent with the
luminosities and colors of LSB knots from Schombert, McGaugh \& Maciel 2013 which
were tentatively identified as star-forming regions based in H$\alpha$ images.
}
\label{clusters}
\end{figure}

Numerous stellar associations are identified in all three galaxies, again indicating
that star formation in LSB galaxies proceeds in a fashion similar to HSB star-forming
spirals and irregulars, i.e. molecular clouds collapsing to form stellar clusters.
Thirty-nine groupings were identified in all three galaxies (23 in F415-3, 11 in
F608-1 and 5 in F750-V1).  Assuming a standard IMF, these clusters range in masses
from a few times $10^4$ to $10^6$ $M_{\sun}$, see Figure \ref{clusters}.  However, as
a cautionary note, most of these groupings are 25 to 50 pc in size and are probably
collections of several smaller clusters in the same region.  Open clusters
found in M31 (Williams \& Hodge 2001) display a similar appearance to the clusters
identified in our LSB galaxies (see their Figure 5), but the regions associated with
H$\alpha$ emission are 100's of pc in diameter and can accommodate several groups of
ionizing OB stars.

The star clusters' mean colors are typically blue ($V-I < 0.5$), but the inclusion of
one or two bright red red giant branch (RGB) or asymptotic giant branch (AGB) star
makes a mean color less sensitive to the age of the cluster (see Asa'd \& Hanson
2012).  Our limiting magnitude (see \S2.2) means that the stellar sources that
identify the clusters are composed solely of OB stars or stars above the tip of the
RGB. The clusters in F415-3 are bluer, on average (mean $V-I = 0.0$), versus the
cluster colors in F608-1 and F750-V1 (mean $V-I = 0.5$).  These colors map
consistently into the $B-V$ of LSB knots from ground-based images (Schombert, McGaugh
\& Maciel 2013) implying that the ground-based colors of the enhanced surface
brightness knots are also driven by the brightest stars.

\begin{figure}[!ht]
\centering
\includegraphics[scale=1.5,angle=0]{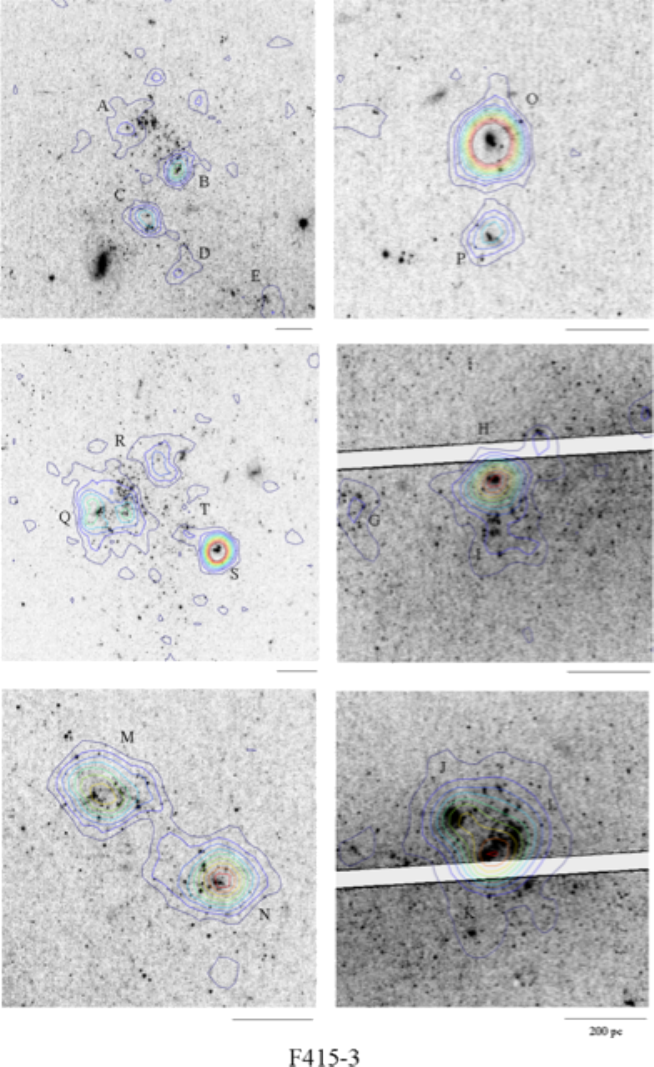}
\caption{\small Selected star clusters from F415-3 F555W images.  Colored contours
display H$\alpha$ emission from ground-based images (FWHM = 1.2 arcsecs).  Clusters A
through E are located in a LSB knot at SW corner of the galaxy.  The knot resolves
into four distinct clusters, all associated with H$\alpha$ emission.  Cluster O
displays a bright HII region powered by a pair of massive O stars ($M_I < -5$).
Cluster Q through T are associated with a LSB knot to the SE.  Clusters M and N
display the more typical shape and size of an LSB star forming region.
}
\label{F415-3_clusters}
\end{figure}

Figures \ref{F415-3_clusters}, \ref{F608-1_clusters} and \ref{F750_clusters}
display a mosaic of some clearer examples of the star associations in each galaxy.
Associations are identified by a letter and H$\alpha$ contours from ground-based imaging
are also shown in each Figure.  The H$\alpha$ peaks are always associated with a
bright, blue star or small grouping of blue stars.  However, there are several
associations not identified with H$\alpha$ emission (e.g., cluster C in F608-1).
Normally, these would be identified with older associations (ages greater than 10
Myrs) as the ionizing stars would have died off; however, most have at least one
centrally located bright, blue star.  Presumingly, these are clusters where the
leftover gas has been blown away by galactic winds or made too diffuse to be detected
in ground-based H$\alpha$ imaging.

F415-3 is our largest galaxy in sample with the most number of detected stellar
sources.  Twenty-three associations were identified by visual inspection, all associated
with a distinct HII region.  The irregular morphology of LSB galaxies is often
driven by the presence of one or two knots, now identified with a single cluster or
group of clusters.  For example, the LSB features to the SE and SW (see Figure
\ref{ground}) are cluster groups A through D (SW) and groups Q through S (SE).
However, their spacing is sufficiently wide enough to prevent a distinct HSB knot as
would be visible in a star-forming irregular galaxy.  Cluster O demonstrates the
sharp correlation between ionizing star luminosity and the H$\alpha$ emission. The
ionizing pair of OB stars in cluster O are the brightest in the sample.

\begin{figure}[!ht]
\centering
\includegraphics[scale=1.5,angle=0]{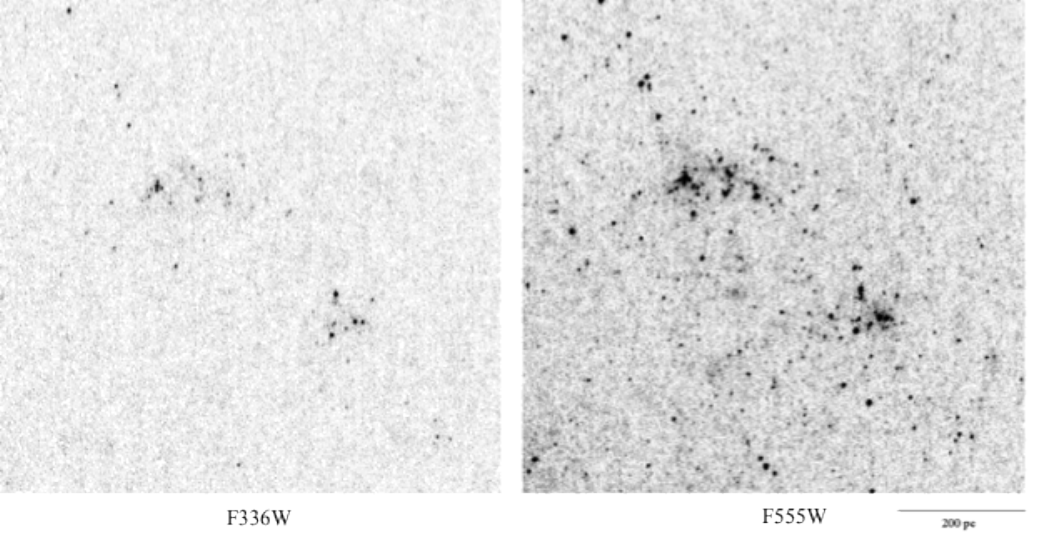}
\caption{\small A comparison of the F336W and F555W images for clusters M and N in
F415-3.  The H$\alpha$ emission in Figure \ref{F415-3_clusters} centers around the
bluest handful of stars in each cluster.  The F336W images are decisive in identifying
the bluest stars in each galaxy.
}
\label{F336_F555_cluster}
\end{figure}

\begin{figure}[!ht]
\centering
\includegraphics[scale=1.5,angle=0]{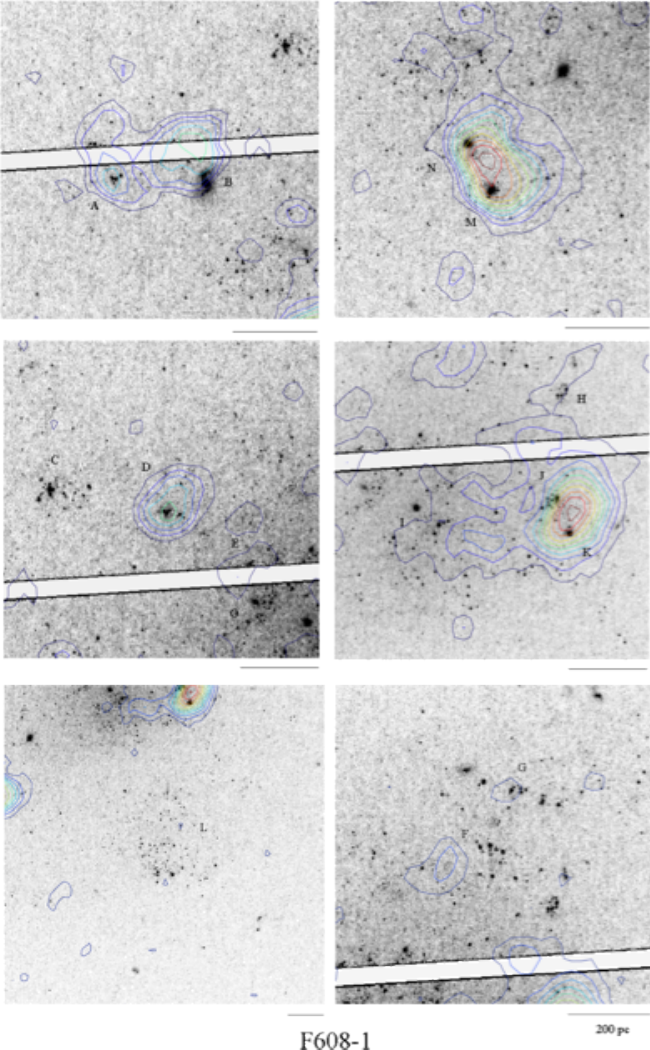}
\caption{\small Selected star clusters from F608-1 F555W images.  Colored contours
display H$\alpha$ emission from ground-based images (FWHM = 1.2 arcsecs).  Unlike
F415-3, several clusters in F608-1 lack H$\alpha$ emission (clusters C, F and L).
Inspection of their localized CMD's displays a lack of bright OB stars with a strong
young RGB population (rHeB stars, see \S2.6) indicating a older population than those
that comprise the H$\alpha$ regions.
}
\label{F608-1_clusters}
\end{figure}

\begin{figure}[!ht]
\centering
\includegraphics[scale=1.5,angle=0]{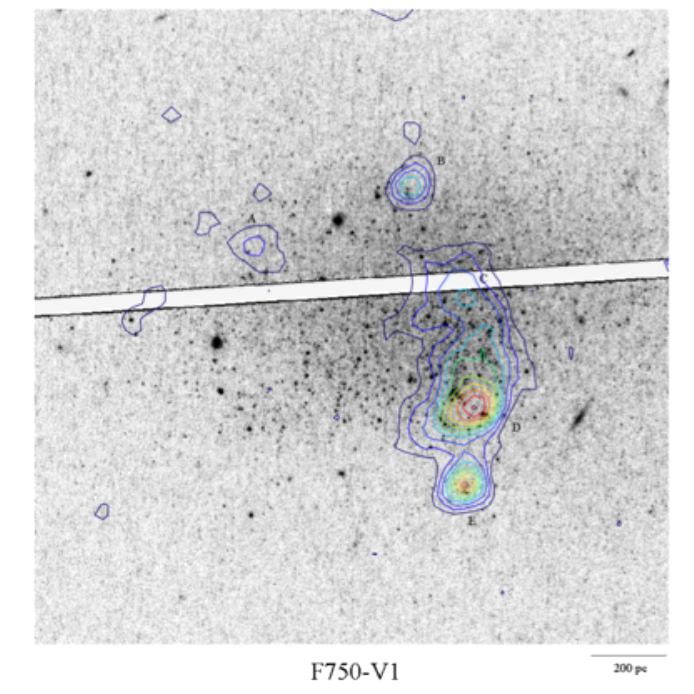}
\caption{\small F750-V1 is the smallest galaxy in the sample with only five HII
regions.  While each HII region is identified with an ionizing OB star, no large
association of stars are visible like those in F415-3 or F608-1.
}
\label{F750_clusters}
\end{figure}

Clusters M and N (bottom left) are good examples of two H$\alpha$ knots associated
with two distinct knots in $V$ ground images.  The two clusters are displayed in
Figure \ref{F336_F555_cluster} with the F336W frame next to the F555W frame.  The
ionizing blue stars are obvious in the F336W frame, with 4 to 5 OB stars per cluster.
This maps nicely into the H$\alpha$ fluxes of those regions (log $L_{H\alpha} =
37.13$ and 37.15 respectfully) which corresponds to approximately five OB stars (see
Figure 13; Schombert, McGaugh \& Maciel 2013).

The central HII region in F415-3 divides into at least three groupings (J, K and L).
The H$\alpha$ flux of this region (38.10) corresponds to several Orion sized HII
regions, but was unresolved into the distinct clusters.  We suspect that many of the
bright HII regions in LSB galaxies are unresolved combinations of several Orion-sized
clusters as displayed by clusters J, K and L as their combined surface brightness is
only slightly less that a 30 Dor sized complex.  Cluster H and I (middle right panel)
display a more common grouping where cluster H is powered by a massive O star, while
cluster I has only two, faint stars visible in the F336W frame and would produce very few
ionizing photons.

F608-1 also displays a number of distinct associations (see Figure \ref{F608-1_clusters},
ranging from $10^4$ to a few times $10^5$ $M_{\sun}$.  Of the thirteen identified
groupings, only two are not associated with H$\alpha$ emission.  One grouping (L) is
widely dispersed, having the appearance of an old, open cluster, but being much too
dispersed to be a gravitational unit (having a scale size of 400 pcs).  Most 
likely, this is a region where several complexes were born, evolved and dispersed by
kinematic effects.  The G cluster displays very faint H$\alpha$ under closer inspection of the
original ground-based images.  The remaining cluster (C, middle left) shows no
H$\alpha$ despite having several OB stars to ionize any nearby gas.  It is one of the
reddest clusters in the sample ($V-I=1.5$) perhaps indicating an older age where the
leftover gas has been blown away by stellar winds.

Our smallest galaxy, F750-V1 (see Figure \ref{F750_clusters}), has five distinct
HII regions but with H$\alpha$ luminosities near log $L_{H\alpha}$ = 36.0, i.e. the
flux expected from one OB star.  Cluster E is a distinct cluster of a handful of blue
stars, but the HII regions to the north are a widely dispersed, over an area of 500
pc, with only a few OB stars.  Clusters A and B are powered by a single O star.

\subsection{Stellar Distribution}

Without a global dynamic structure, such as supplied by spiral density waves, the
star formation history in LSB galaxies should be dominated by stochastic processes
and, thus, the spatial distribution of young stars is a measure of this process.  As
can be seen in Figure \ref{map1}, the blue stars are clearly more clustered then the
redder population, although this is even more magnified as many of the $V-I>0.5$
stars are on the red helium-burning branch and, thus, are only 100 Myrs old.

This differs from the stellar distribution in LV dwarfs such as NGC 1705 (Tosi \etal
2001) where the young stars are concentrated in the central regions with an older
population found in the halo.  However, the distribution of young stars in
our irregular LSB galaxies is mostly a statement of how the stellar mass is
distributed.  Most irregular LSB galaxies have no well-defined central location and
it is rare to find the highest surface brightness region associated with the geometric
center defined by the outer isophotes.  Star formation, and thus the youngest stars,
are clearly associated with the higher surface brightness knots seen in the
ground-based images and the uniform stellar distributions, such as those in NGC 1705, are
accidents of the uniformity of the isophotes in some LV dwarfs.

To be more precise, 75\% of the stars in identified stellar associations or groupings have
colors less than $V-I=0.5$ while only 40\% of the field population are that blue.
Although we refer to the groupings in Figure \ref{F415-3_clusters},
\ref{F608-1_clusters} and \ref{F750_clusters} as star clusters, this is a misnomer
as gravitational bound open clusters range have sizes from a few to 10 pc's in diameter.
In fact, these associations should be referred to star complexes, for they probably
contain several cluster-sized units and their luminosity plus H$\alpha$ fluxes are
more in agreement with a grouping of young open clusters.

The star formation pattern in our LSB sample is similar to the pattern found in
Sextans A (a Local Group dwarf of similar size and luminosity as F415-3, see
Dohm-Palmer \etal 2002) where the brightest bMS stars are found in the stellar
groupings and the highest percentage of older AGB stars are found in the regions
between the H$\alpha$ knots.  Also the 100 Myr population (rHeB stars) are located
primarily in the stellar associations, but with less density than the very young bMS
stars (50\% versus 75\%).  None of this is particularly surprising as it appears that
local star formation in LSB galaxies proceeds in the same fashion as their HSB
irregular cousins, i.e., from compact clusters to dispersed associations.

The only peculiar aspect to the stellar distribution (especially for F415-3) is the
existence of any bright O stars outside of an association or an HII region.  The lack
of H$\alpha$ emission may relate to the low gas density in LSB galaxies for about
50\% of the H$\alpha$ emission in LSB galaxies is not associated with a particular
HII region, but exists in a low surface brightness diffuse form.  Thus, the isolated
O stars may be generating this H$\alpha$ emission not visible against the sky
background.  O stars outside of a cluster is not improbable considering the internal
kinematics of LSB galaxies.  The typical gas velocity dispersion is 8 km/sec (Kuzio
de Naray \etal 2006) which corresponds to 5 pc/Myr, or sufficient velocity to scatter
older O stars from their regions of intense star formation, or isolate single O star
HII regions.

\subsection{Color Magnitude Diagrams}

\begin{figure}[!ht]
\centering
\includegraphics[scale=0.9,angle=0]{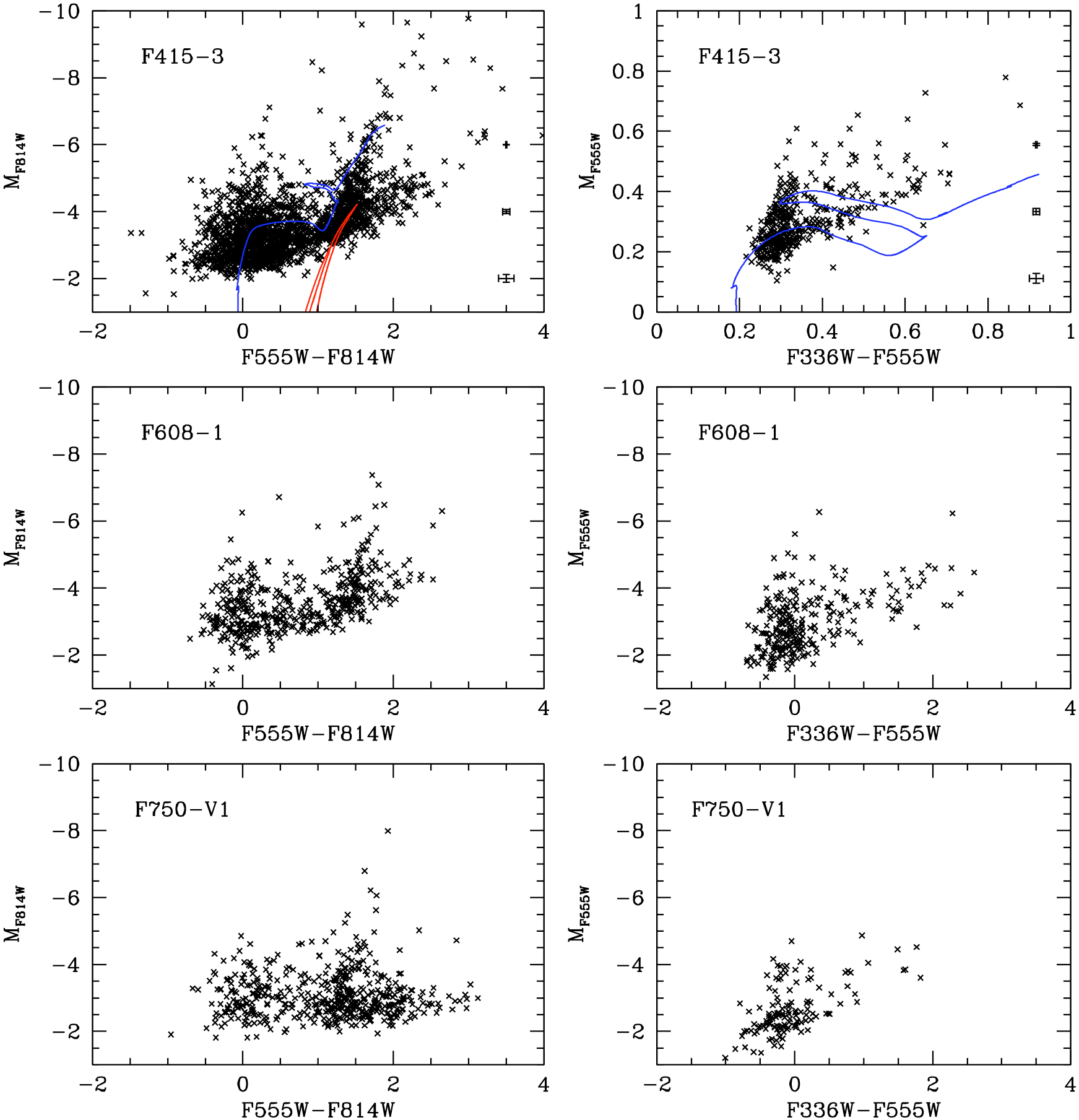}
\caption{\small The two color CMD's for all three LSB galaxies are displayed with no
extinction corrections, but using the distance moduli in Table 1.  Similar CMD
features are seen in each galaxy with prominent blue main sequences and young RGB's
(rHeB populations, see discussion in text).
Low metallicity isochrones for a 50 Myr and 12 Gyr population are shown.  Typical
error bars are shown on the right side of the top panels.
}
\label{cmd}
\end{figure}

The color-magnitude diagrams for all three galaxies, in HST filters F336W, F555W and
F814W, are shown in Figure \ref{cmd}.  The left panels are $M_{F814W}$ versus
$F555W-F814W$, the right panels display $M_{F555W}$ versus $F336W-F555W$, where the
absolute magnitudes are determined from the distances in Table 1.  Shown for
reference are stellar isochrones for a 50 Myr population of [Fe/H]=$-$0.6 and a 12
Gyr population with a [Fe/H]=$-$1.5.  Similar features are seen in all three
galaxies, with the clearest CMD morphology visible in F415-3 due to its larger
sample.

The brightest stars in F415-3 ($M_{F814W} > -8$) are potentially foreground stars.
According to Gould, Bahcall \& Maoz (1993), a total of 3$\times$10$^5$ stars per
square degree are expected at this galactic latitude and limiting magnitude of
$M_V=-7$ ($m_V=23$).  For the size of F415-3, this results in 15 possible
contaminating stars, whereas there are 16 stars brighter than $M_{F814W} = -8$ in
F415-3. This makes all of them potentially non-members of the F415-3's stellar
population.  Neither F608-1 nor F750-V1 appear to have foreground stars contaminating
their samples, probably due to their smaller samples and angular sizes.

Our analysis of the CMD's in LSB galaxies is guided by comparison to other dwarf
CMD's (e.g., IC 2574, see below) and stellar population simulations.  One of the
highest quality simulators is the synthetic CMD generator from IAC-STAR (Aparicio \&
Gallart 2004).  Numerous variables control a synthetic CMD generator, i.e., the IMF,
star formation rates and chemical evolution scenarios.  Our experiments used the
isochrones (Bertelli \etal 1994) with the default mass loss and IMF settings from the
IAC-STAR simulations.  We adopt a chemical evolution scenario similar to our own
stellar population models (Schombert \& McGaugh 2014a) where the initial metallicity
was chosen to be [Fe/H] = $-$1.5 and ending with values varying from $-$0.9 to +0.1.

\begin{figure}[!ht]
\centering
\includegraphics[scale=1.7,angle=0]{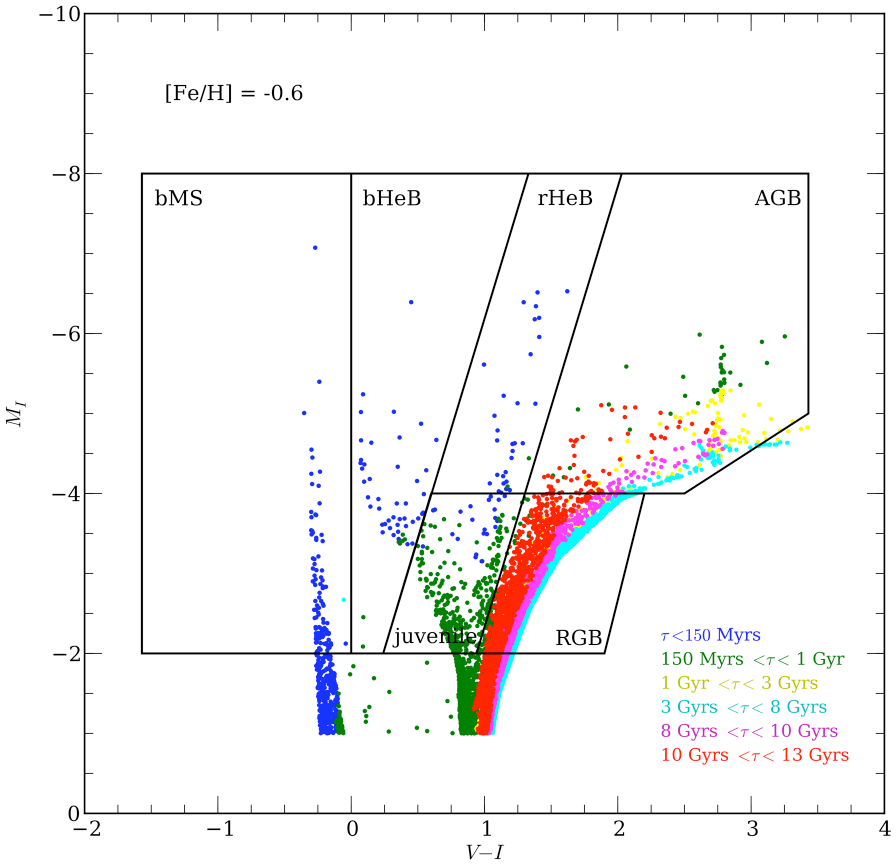}
\caption{\small An IAC-STAR synthetic CMD for a stellar population of 13
Gyrs in age with a constant star formation rate.  The starting population has a
metallicity of Z=0.0006 ([Fe/H]=$-$1.5) and a final metallicity of Z=0.004
([Fe/H]=$-$0.6).  Various colors correspond to different ages; blue = less than 150
Myrs, green = 150 Myrs to 1 Gyr, yellow = 1 to 3 Gyrs, cyan = 3 to 8 Gyrs, magenta =
8 to 10 Gyrs, red = greater than 10 Gyrs.  Also shown are the CMD morphology regions
used in Figure \ref{ic2574}.  The youngest population ($\tau < 150$ Myrs) is
displayed in greater detail in Figure \ref{age_color_mag}.
}
\label{iac_regions}
\end{figure}

One such synthetic CMD's, for a final metallicity of [Fe/H]$=-0.6$ and constant star
formation over 13 Gyrs, is shown in Figure \ref{iac_regions}.  Only the top of the
CMD is shown ($M_I < -1$) and different aged stars are represented with different
symbol colors (blue for less than 150 Myrs, green for between 150 Myrs and 1 Gyrs,
yellow for 1 to 3 Gyrs, cyan for 3 to 8 Gyrs, magenta for 8 to 10 Gyrs and red for
older than 10 Gyrs).  Immediately obvious are the very young features of a stellar
population, the bMS, bHeB and rHeB branches plus a fainter 'juvenile' population of
stars with ages between 150 Myrs and one Gyr (this would include the brightest
portion of the red clump).  Beyond one Gyr, the stellar populations quickly develops
into a classic RGB with the oldest stars forming the blue edge of the RGB.
Intermediate aged stars (between two and eight Gyrs) dominate the AGB region of the
synthetic CMD.

We have designated specific regions in the $M_I$ vs $V-I$ CMD to compare the star
formation history of LSB galaxies with the other HST dwarf galaxy samples.  This will
allow us to compare data from dwarf galaxies with large numbers of stellar sources to
our smaller samples plus comparison with synthetic CMDs.  The six regions are shown in
Figure \ref{iac_regions} and comprise the area in the CMD that are sensitive to specific
age and metallicity effects.  The region to the far blue encompasses the youngest
stars, those making up the tip of the main sequence (bMS), defined as all stars bluer
than $V-I=0.0$.  Slightly redder is the region containing the blue branch of the
helium-burning phase (bHeB) defined by a wedge parallel to the red (rHeB) branch.
The rHeB branch is defined as a parallelogram with a width that would capture a range
of metallicities strictly above the $M_I = -4$ line, i.e. the tip of the RGB.

Stars in bMS region are less than 15 Myrs in age, while stars in the bHeB and rHeB
are between 15 and 150 Myrs. The rHeB feature is from $M_I < -4$ and between $1 < V-I <
2$ and, while this region also contains very young stars, the age of the stars
decreases as their luminosity increases (see Figure \ref{age_color_mag}).  Therefore,
the number of the stars from the base of the rHeB to the tip are a measure of star
formation over the last 100 Myrs (see \S2.8).  A combination of studying the
bluest stars, and the brightest red stars, resolves the most recent star formation
epoch outside of H$\alpha$ emission ($\tau < 15$ Myrs).

\begin{figure}[!ht]
\centering
\includegraphics[scale=1.9,angle=0]{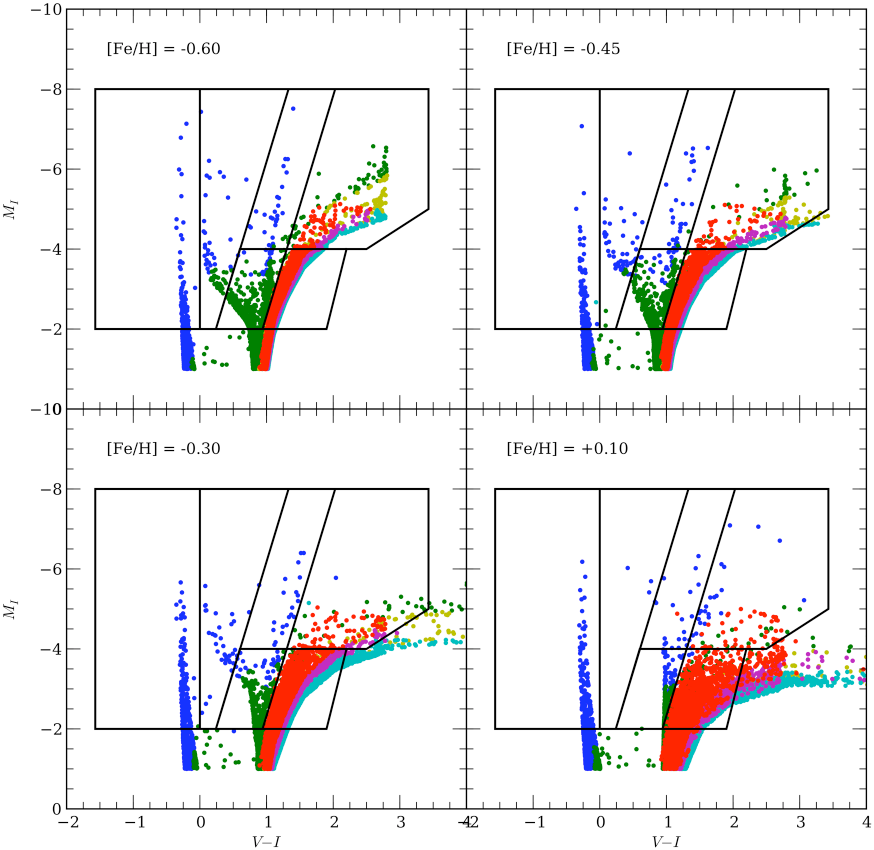}
\caption{\small Four IAC-STAR simulations for ending values of [Fe/H] of $-$0.60,
$-$0.45, $-$0.30 and +0.10.  The rHeB branch is well defined for low metallicities
and degrades with higher values, as well as drifting to the red.  The population of
AGB stars is dominated by intermediate age stars for low metallicities, but
decreasing in numbers at higher metallicities.  The population regions outlined in
Figure \ref{iac_regions} are marked.
}
\label{iac}
\end{figure}

The region below the rHeB contains stars with ages between 150 Myrs and 1 Gyrs, a
young population, but not the stars involved in HII regions or any emission line
signatures of star formation.  As they are younger that the typical intermediate age
population (e.g., AGB stars), we have titled this region as 'juvenile' stars.  
Stars older than one Gyr will occupy the AGB and RGB sections of the CMD.  The stars
with ages between one and 8 Gyrs dominate the AGB region, thus the ratio of AGB to
RGB stars is a measure of this epoch of star formation.  However, this region is also
highly dependent on the metallicity where lower metallicity populations contain more
stars in the AGB region.  The effect of metallicity can be seen in Figure \ref{iac},
where four simulations of varying ending [Fe/H] are shown.  Metallicity
effects are most prominent for the rHeB and AGB populations.

Lastly, there is the region below $M_I = -4$ (the tip of the RGB) that is the classic
old, RGB.  The blueward side of the old RGB is fixed by the metallicity of the
initial stellar population.  As the metallicity, and age, increases for later
generations, those stars occupy the redder portion of the CMD.  Low metallicity, old
stars can occupy the AGB region but with decreasing number of AGB's with increasing
metallicity.  Thus, the ratio of the AGB region to the old RGB region, combined with
the position of the rHeB, is a measure of the rate of chemical evolution of a galaxy
and its current metallicity.

Interpretation of the various regions is dependent on the chemical history of the
galaxy.  To demonstrate this effect, we varied the final metallicities between
[Fe/H]$=-0.6$ and +0.1 for a population with a history of constant star formation and
a initial metallicity of [Fe/H]=$-$1.5.  The results are shown in Figure \ref{iac}
for the four different ending [Fe/H] values.  The first characteristic to note is
that the position of very young stars (less than 15 Myrs, the bMS) is independent of
metallicity.  On the other hand, the color of the bHeB and rHeB branches are mildly
dependent on metallicity.  In fact, the rHeB is a feature that only exists for
metallicities less than $-$0.3 and its mean color is sharply defined by the current
metallicity.

Other features to note is the broadening of the RGB with increasing metallicity, and
the decreasing importance of AGB stars with metallicity.  For the range of expected
metallicities (LSBs range from [FE/H] = $-$1.0 to $-$0.3 based on oxygen abundance
analysis, McGaugh 1994; Kuzio de Naray, McGaugh \& de Blok 2004), the morphology of
the RGB and AGB are fairly constant.  The juvenile population also remains well
defined over these metallicities, with the dividing line between the juvenile and RGB
populations unchanged by variation in metallicity.  Changes in star formation history
will work to increase (or decrease) the numerical proportions of the various ages,
but will not alter the position of the various regions in color-luminosity space.

A complicating factor to the simulations is the ratio of so-called light or $\alpha$
elements, typically expressed as $\alpha$/Fe, in a stellar population.  Supernovae
are the main contributors of metallicity in a stellar population; however, Type Ia
and Type II SN contribute differing amounts of Fe where Type Ia SN overproduce Fe
with respect to $\alpha$ elements.  Thus, the ratio of $\alpha$ elements to Fe is a
function of the number of Type II versus Type Ia supernova in a galaxy's past with
Type Ia SN producing extra amounts of Fe and driving the ratio downward.  Since main
source of free electrons in stellar atmospheres is Fe, stars with high $\alpha$/Fe
ratios will have hotter (i.e., bluer) colors.  Ellipticals tend to have high
$\alpha$/Fe ratios (typically near 0.3, compared to the solar value of 0.0,
S{\'a}nchez-Bl{\'a}zquez \etal 2006), primarily due to their evolution dominated by a
rapid burst of star formation at early epochs.  As Type Ia SN require at least a Gyr
to develop their white dwarf companions, a rapid burst of star formation will leave a
stellar population deficient in Fe (i.e., a high $\alpha$/Fe ratio).  However, Type
Ia SN require at least a Gyr of time to build-up within a galaxy and, therefore, high
fractions of $\alpha$/Fe indicate shorter duration times for star formation.
Constant star formation, such as found for the Milky Way, allows for the build-up of
Type Ia SN and, therefore, lower $\alpha$/Fe values.

We investigated the effects of variation in $\alpha$/Fe in Paper III (Schombert \&
McGaugh 2014a) using $\alpha$-enhanced isochrones from the BaSTI group (Pietrinferni
\etal 2004).  Increasing the $\alpha$/Fe ratio by 0.3 from solar resulting in
integrated $B-V$ color of 0.03 bluer.  The $V-I$ isochrones are approximately 0.02
bluer around the red clump, decreasing to zero at the turn-off point.  Similar
changes were observed for the bHeB and rHeB positions in the $\alpha$-enhanced
isochrones.  The expectation for the star formation history of LV and LSB dwarfs is
that their past SFR's are more similar to the Milky Way than ellipticals.  So
expected values for $\alpha$/Fe should range from solar (i.e. zero) to slightly less
than zero.  For example, Lapenna \etal (2012) found the youngest stars in the LMC to
be slightly under solar ($\alpha$/Fe=$-$0.1).  If the stars in the LSB galaxies in
are sample have similar ratios, than the effect of $\alpha$/Fe on the synthetic CMD's
will be quite small (less than 0.01 in $V-I$) and comparison with LV dwarfs (with
near solar values) is appropriate.

\begin{figure}[!ht]
\centering
\includegraphics[scale=1.0,angle=0]{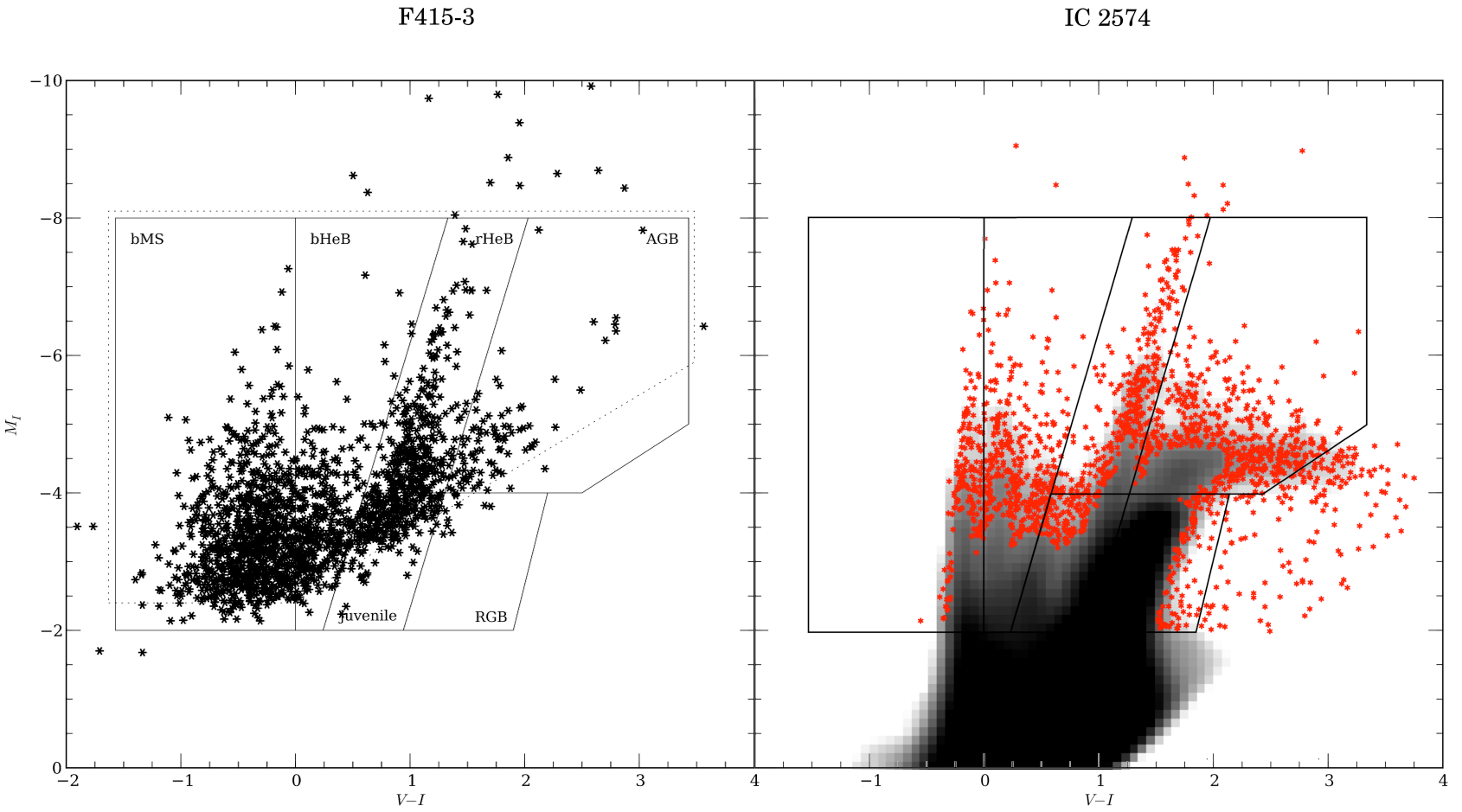}
\caption{\small A comparison of the CMD's for F415-3 and IC 2574 from Dalcanton \etal
(2009).  Several CMD morphology features are evident in both galaxies.  Particular
regions of interest are marked; the blue main sequence (bMS), the blue and red helium-burning branches (bHeB and rHeB), a "juvenile" population (between 100 Myrs and 1
Gyr), the AGB and RGB populations ($\tau > 3$ Gyrs).  These regions are defined by
comparison to synthetic CMD simulations (see Figure \ref{iac_regions}).  The dotted
line displays the completeness limit used when comparing fractions of the various CMD
features with other CMD's.  IC 2574 is displayed with a logarithmic Hess diagram and
red symbols for stars in regions of the CMD with few stars.
}
\label{ic2574}
\end{figure}

In order to compare the deeper CMD's of nearby dwarfs to our three LSB galaxies, we
have outlined a completeness region (see Figure \ref{ic2574}) that includes 95\% of
the stars in our target galaxies.  The same completeness region is applied to the
CMD's taken from the Extragalactic Distance Database (EDD, Jacobs \etal 2009) and to
the synthetic CMD's generated from IAC-STAR.  Thus, it is important to remember that
the fractional values quoted in the following discussions do not represent the
percentages with respect to the entire galaxy stellar population.  They only refer to
the completeness region although an extrapolation to the total population could be
made with some simple assumptions to the distribution of faint stars.

While the depth of the CMD's for our three LSB galaxies does not approach the depth
of other CMD studies in galaxies (i.e., a main sequence and turnoff, Jacobs \etal
2009), a clear red helium-burning sequence (rHeB) is visible as well as the top of
the blue main sequence (bMS, also called the blue plume) along with a significant
intermediate age AGB population (see Figure \ref{ic2574}).  The rHeB branch is of
particular interest for it only develops in metal-poor populations ([Fe/H] $< -0.3$)
and the number of stars along the branch is a direct measure of the star formation
rate for the last 100 Myrs (see \S2.8).  The stars in the bMS are consistent with a
young, high mass OB star population, the stars responsible for the low level
H$\alpha$ emission in LSB galaxies.  All the stars found in F814W ($I$) are detected
in F555W ($V$) with the exception of five stars.  These five stars are detected in
F814W between $M_I=-4$ and $-$6, but not visible in F555W.  This predicts their colors
are greater than 2, typical of extreme AGB stars.  The bHeB population is
poorly defined in $V-I$, but is clearer in $U-V$ (see \S2.9).

The LSB CMD's mostly closely resemble the CMD's of LV starburst dwarfs from McQuinn
\etal (2010) and the ANGST survey (Dalcanton \etal 2009) (comparison CMD's can be
found at the Extragalactic Distance Database, Jacobs \etal 2009).  In particular, the
morphology of the CMD in our LSB sample closely resembles the morphology of the CMD
from IC 2574 (McQuinn \etal 2010), one of the faintest (and lowest metallicity) of
their sample.  A comparison between F415-3 and IC 2574 CMD's is seen in Figure
\ref{ic2574} where the IC 2574 data contains 158,000 stars and, thus, is displayed
using a logarithmic Hess diagram overlayed with individual datapoints in the CMD regions
of low density.  Regions of particular interest in the star formation
history are marked.  As can be seen in Figure \ref{ic2574}, the dwarfs in the EDD
catalog display all the CMD features for stellar populations with a range of ages,
such as a blue main sequence, an old red giant branch (divided by the tip of the RGB
at $M_I=-4$), blue and red helium-burning sequences (bHeB and rHeB), a red clump (RC)
and an asymptotic giant branch (AGB).  Features in common with F415-3 are a distinct
rHeB branch and AGB population.  A bMS is evident in both galaxies, but that feature
in F415-3 is broader due to increased photometric errors near the completion limit.
We note that the LSB galaxies in our sample have very little resolution of the old
RGB populations.  Several bMS tracks are visible in IC 2574, indicating bursts of
star formation on timescales of 10 Myrs.  Similar features are not seen in F415-3,
probably due to small number statistics.

\subsection{CMD Morphology}

Using the CMD regions defined in Figure \ref{iac_regions}, we can classify the CMD
morphologies of existing HST samples from the EDD (Jacobs \etal 2009) for comparison
with our LSB galaxies.  We have divided the existing CMD samples from EDD into young
(CMD's with a clear rHeB branches and strong AGB populations, such as IC 2574)
and old (ones lacking a rHeB branch, but may have a weak bMS populations).  Examples
of old morphologies are DDO 44, DDO 71 and ESO 294-010 from the ANGST survey.  In
all, 57 CMD's were extracted from the HST archives and the EDD website, 45 classified
as young and 12 classified as old.

\begin{figure}[!ht]
\centering
\includegraphics[scale=0.8,angle=0]{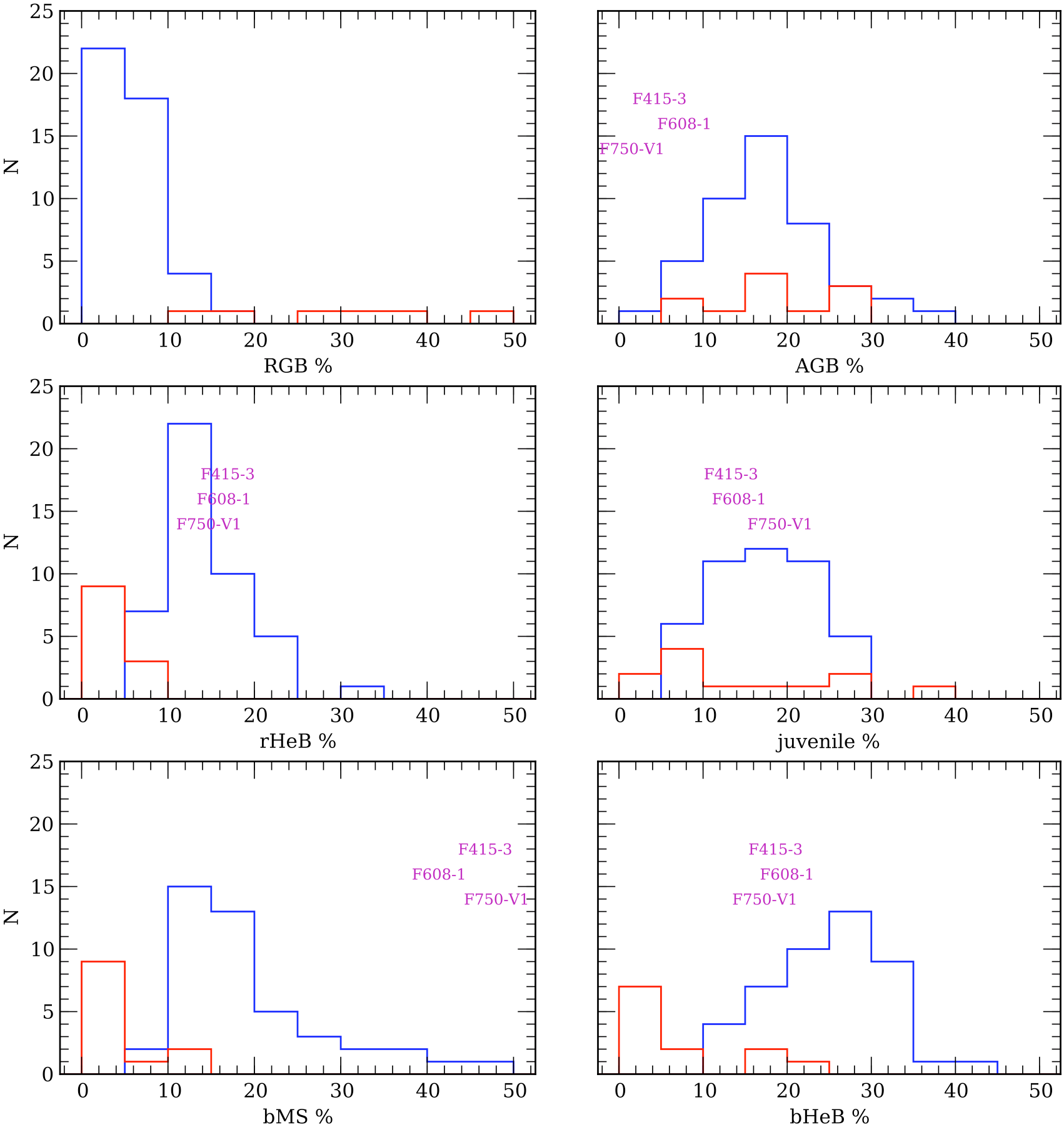}
\caption{\small A comparison of the fraction of stars in the six population regions
defined in Figure \ref{iac_regions}.  In general, young (blue) and old (red) CMD's separate based
on the dominance of blue plume features (bMS, bHeB and rHeB) versus RGB fractions.
Our three LSB galaxies are indicated and differ from young dwarfs by having stronger
bMS and weaker AGB fractions.  Incompleteness prevents any strong statements on the RGB
population in the LSB galaxies.
}
\label{fraction_hist}
\end{figure}

Each CMD of the three LSB galaxies in our sample, is analyzed by calculating the
number of stars in the six population regions outlined in Figure \ref{iac_regions}.
The population percentages are displayed as histogram in Figure \ref{fraction_hist}.
The percentage of bMS stars varied from only a few percent for old dwarfs (e.g. DDO
44) to over 40\% for dwarfs such as NGC 3077 and UGC 5336.  The galaxies with strong
rHeB branches have bMS populations that vary between 10 and 40\%, indicating a
connection between the two features but the bMS being more sensitive to very recent
SF.  For example, the fraction of stars in the rHeB branch ranges between 10 and 20\%
of the population, regardless of the fraction of bMS stars, due to the fact that the
bMS fraction varies on very short timescales.  The old dwarfs in the EDD sample
display strong RGB fractions and weak bMS and rHeB branches.  Both young and old EDD
dwarfs have similar AGB fractions (suggesting their primary differences is due to
their star formation rates over the last Gyr).  The galaxies without prominent rHeB
branches display the highest concentrations of RGB stars in the completeness region,
reinforcing the interpretation that old dwarfs, while often having some current star
formation, produced most of their stars over 5 Gyrs ago.

Young dwarfs typical have strong bMS, bHeB and rHeB populations (McQuinn \etal 2010)
which would agree with the fact that their current SFR (based on H$\alpha$ values)
exceeds the mean past SFR based on dividing their stellar mass by 12 Gyrs (i.e.,
$<$SFR$>$, see \S3.2).  Increasing importance of the bMS stars in young dwarf CMD's
reflects into decreasing percentages of RGB stars, i.e., star formation has continued
to recent epochs.  The constant fraction of rHeB stars indicates that the bursts of
star formation responsible for the bMS population are fairly evenly spacing on
timescales of 100 to 200 Myrs.

The three LSB galaxies in our sample distinguish themselves from the young dwarfs by
having weaker AGB fractions and stronger bMS fractions.  The LSB galaxies also have
weaker juvenile and RGB fractions; however, the RGB region is undersampled and, even
when we apply the same completeness boundaries to all the CMDs, we are hesitant to draw
strong conclusions from this trend.  The LSB galaxies have similar bHeB and rHeB
branch fractions with the star-forming EDD dwarfs, which samples the Gyr timescale of
star formation.

\begin{figure}[!ht]
\centering
\includegraphics[scale=0.8,angle=0]{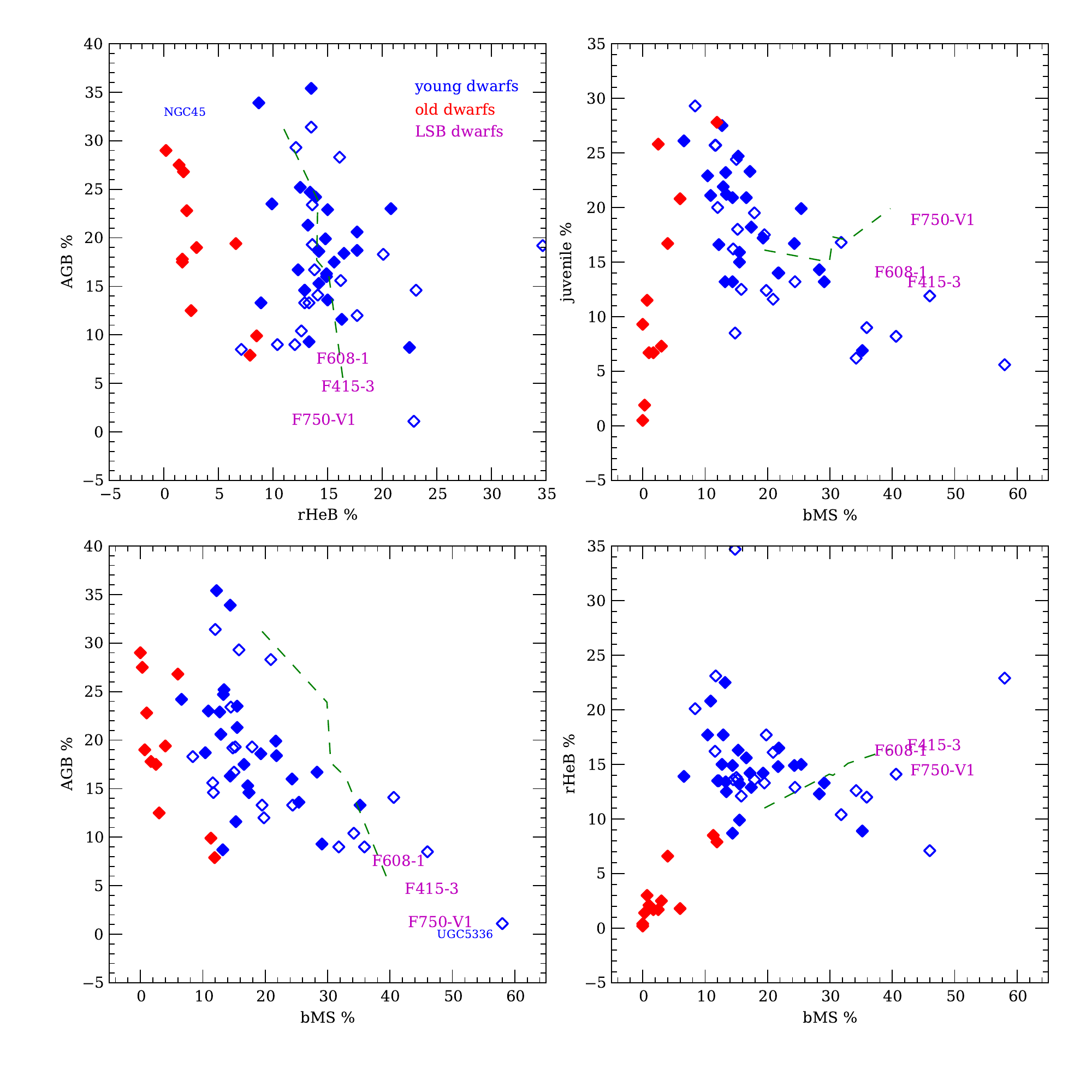}
\caption{\small The relationships between population fractions for the bMS, rHeB,
juvenile and AGB portions of the CMD.  Low metallicity young dwarfs are shown as open
symbols, high metallicity dwarfs as solid symbols.  LSB galaxies are indicated by
their names.  Old dwarfs are identified by their absence of a distinct rHeB branch.
Also shown are synthetic CMD simulations with increasing SFR's (dashed green line).
}
\label{fractions}
\end{figure}

The difference in young and old populations can be seen more clearly in Figure
\ref{fractions}, a comparison of the fraction of bMS, rHeB, juvenile and AGB stars.
Old dwarves were selected by an absence of a distinct rHeB branch, so their low
values are unsurprising.  They typically, also, have very weak bMS, bHeB and rHeB
populations, signaling very low rates of star formation over the last Gyr.  Old
dwarfs display a range of juvenile and AGB fractions (anti-correlated), suggesting a
continuum driven by a star formation history of increasing SFR from the young to old
dwarfs.

Even though AGB's are a measure of intermediate age stars, there is a strong
anti-correlation between the bMS and the AGB fraction (bottom left panel of Figure
\ref{fractions}.   We see the dominance of AGB stars in the older CMD's, but the
trend of decreasing AGB populations with increasing bMS populations is also evident
in the young CMD's.  As the AGB stars sample intermediate timescales (3 to 8
Gyrs), then we, again, see a trend of increasing star formation from intermediate
ages (although strict interpretation requires comparison to synthetic CMDs, see
\S2.6).  The ratio of AGB to RGB stars (not shown) increases with larger AGB
populations to a maximum of approximately 20\%.  The linear behavior of the AGB to
RGB relation for young dwarfs may signal a late initial star formation epoch, in
agreement with their higher current SFRs compared to their past mean rates.

All three LSB galaxies in our sample have CMD morphologies at the extreme edges of
other dwarf CMD's.  Our LSB galaxies CMD's typically have stronger very young components (i.e.,
bMS) with a mean of 45\% compared to the LV dwarf mean value of 20\%.  The dominance
of the bMS population comes as no surprise due to the extremely blue colors for most
LSB galaxies.  And while their HSB cousins also have blue star-forming colors, those
colors are typically restricted to the bright star-forming regions.  LSB galaxies are
unique in that even the regions between the few higher surface brightness knots are
blue in optical colors (Schombert, Maciel \& McGaugh 2011).  The widely dispersed, and
predominately blue, stellar population are responsible for this effect.

LSB galaxies also have weaker intermediate aged components, with the AGB fraction at
5\% compared to the LV dwarf values between 10 and 30\%.  The interpretation here is
that the current SFR is much higher than the SFR over the last 5 Gyrs, although this
is not supported by the mean past SFR ($<$SFR$>$) as estimated by the current stellar
mass divided by a Hubble time.  The AGB fraction is metallicity dependent (see Figure
\ref{iac}), however, for the estimated [Fe/H] of our LSB sample (between $-$1.0 and
$-$0.6) the AGB fraction should be greater than the higher metallicity LV dwarfs.

The RGB population in LSB galaxies also deficient compared to old LV dwarfs, but
young LV dwarfs have RGB fractions below 10\% so this comparison is difficult.  Also,
the conclusions concerning the $\tau > 8$ Gyrs populations in our LSB galaxies are
less secure due to the lack of complete resolution of the old RGB in our CMD's.  We
deduce the crude characteristics of old stars in LSB's based on this limited
resolution plus the fact that the chemical evolution requires some older population
in order to produce even the low [Fe/H] values measured with the rHeB populations
(see \S2.8 and Villegas \etal 2008).  In addition, the pixel-by-pixel surface surface brightness
characteristics (see \S2.3) also match the expectations for an underlying normal older
population.

Given these limitations, the three LSB galaxies still have very low AGB fractions.
Their rHeB branch fractions are similar to other young dwarfs (although all
star-forming dwarfs have rHeB fractions near 10\%).  A more telling diagnostic is the
ratio of AGB to bMS for LSB galaxies.  While most young, blue dwarfs have an AGB
population in proportion to their bMS populations, the three LSB galaxies have
significantly higher bMS and bHeB populations compared to the AGB populations.  This
is particularly significant since the fraction of AGB's increases with decreasing
metallicity (based on comparison to synthetic CMD's) and both LSB and young dwarfs
display the opposite trend.  Either the star formation rate in LSB's has suddenly
increased in the last Gyr (i.e. the current epoch is a special time) or the past star
formation rates of LSB have long been inhibited, perhaps an explanation for their LSB
nature as a whole.

\subsection{Recent SF and [Fe/H] from the rHeB Branch}

\begin{figure}[!ht]
\centering
\includegraphics[scale=0.8,angle=0]{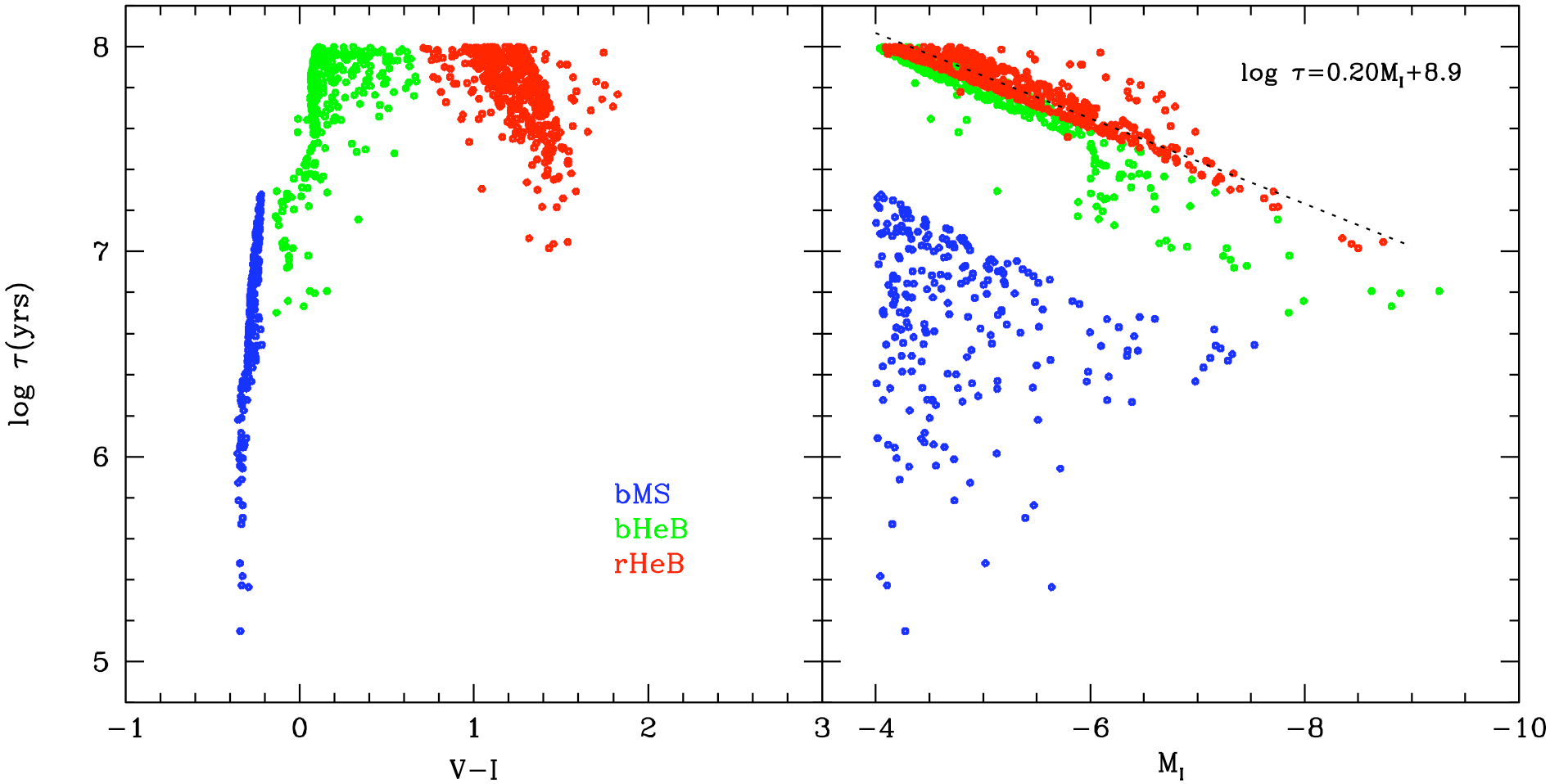}
\caption{\small A synthetic CMD (IAC-STAR) for a metal-poor stellar population
([Fe/H] = $-$0.4) displaying the color and luminosity for the youngest stars (ages
less than 100 Myrs).  The bMS contains only stars less than 15 Myrs and can be easily
distinguished by a simple cut in color above $M_I = -4$.  Older stars occupy the bHeB
and rHeB branches (oscillating between the branches).  The rHeB branch is easier to identify
in the CMD, and the relationship between age and luminosity is linear for stars in the
rHeB region of a CMD (dotted line).
}
\label{age_color_mag}
\end{figure}

The mapping of the most recent star formation has the advantage that the youngest
stars have positions on the CMD that are most easily distinguished from other aged
populations.  In particular, stars with ages less than 100 Myrs are the domain of the
bMS, bHeB and rHeB branches (McQuinn \etal 2012).  Figure \ref{age_color_mag}
displays a breakdown of these three young phases of the CMD from a IAC-STAR
simulation by age, color and luminosity.  Note that a cut by $V-I < 0$ will isolate
all stars younger than $10^7$ years old.  The timescale between $10^7$ and $10^8$
years is represented by the bHeB and rHeB branches.  A star will oscillate between
the two branches, with the younger stars at higher luminosities.  However, the rHeB
branch is much easier to distinguish from the bHeB branch due to confusion in the CMD
between bMS and bHeB stars (however; these two branches are separated in $U-V$ color,
see \S2.9) . In addition, stars on the rHeB branch display a strong correlation
between age and luminosity (see right panel, Figure \ref{age_color_mag}).  This
provides a simple, and direct, measurement of star formation between $10^7$ and
$10^8$.

\begin{figure}[!ht]
\centering
\includegraphics[scale=1.0,angle=0]{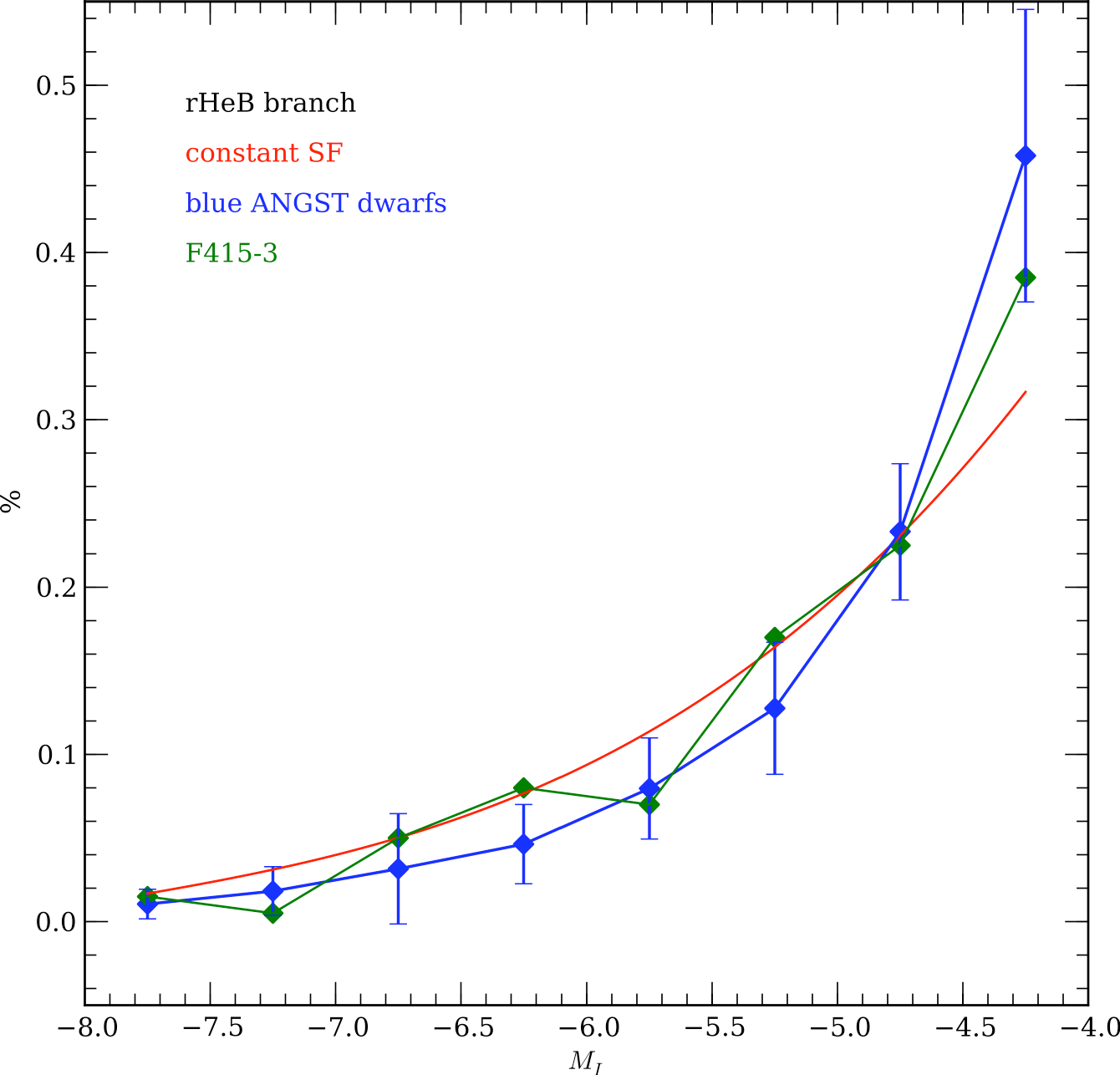}
\caption{\small The distribution of stars along the rHeB branch as a function of
luminosity.  Luminosity along the rHeB branch correlates with age (the youngest stars
being the brightest).  Thus, the number of stars per luminosity bin is a measure of
the star formation history over the last 100 Myrs.  The mean distribution for all
young EDD dwarfs is shown as the blue curve.  A model of constant star formation is 
shown as the red curve.  Nearby dwarfs appear to have slightly higher than constant
SF at 100 Myrs, dropping below the constant curve in recent epochs.  The LSB F415-3
is also shown (green curve), which displays the opposite trend from the EDD dwarfs.
}
\label{rheb}
\end{figure}

The linear relation between absolute luminosity and age for rHeB branch stars (see
Figure \ref{age_color_mag}) allows the distribution of rHeB branch stars on the CMD
to be compared with various star formation histories.  For example, in Figure
\ref{rheb}, the averaged distribution of blue LV dwarfs is compared to an IAC-STAR
synthetic CMD using a constant SFR (the results were independent of metallicity).
The mean star formation history for young dwarfs is slightly higher for ages greater
than 50 Myrs and slightly lower for younger populations, although the range for all
LV dwarfs is consistent with approximately constant SF for the last 200 Myrs.

The SF history of F415-3 (our LSB with the best sampling of the rHeB branch) is
similar to other young LV dwarfs.  F415-3 has a slightly lower SFR at 100 Myrs,
rising to a constant SF by 50 Myrs.  However, F415-3 is well within the distribution
of SF histories of other young dwarfs, again, surprisingly considering the very
different appearance of LSB dwarfs and LV dwarfs in terms of mean stellar density.
If both types of galaxies have similar current SFR, then their differences lie
in their intermediate and older populations, i.e., the mean past SFRs.

\begin{figure}[!ht]
\centering
\includegraphics[scale=1.0,angle=0]{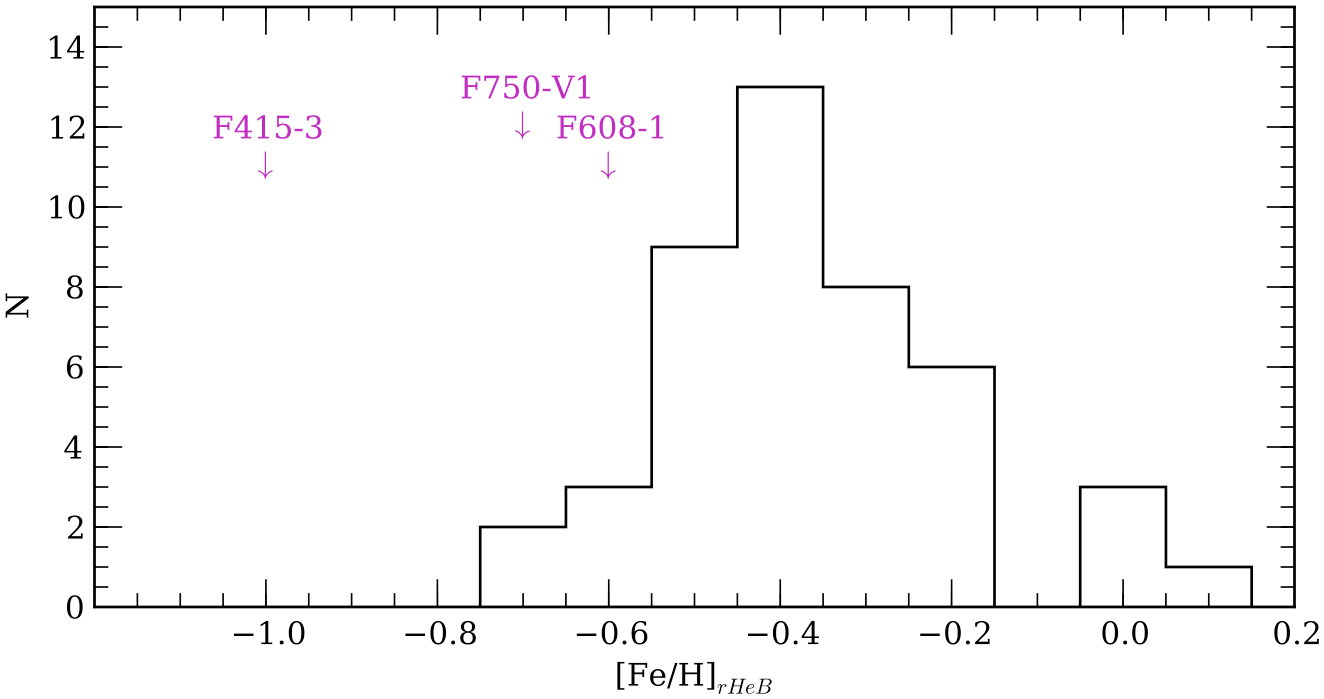}
\caption{\small The distribution of metallicity (parameterized as [Fe/H]) deduced
from the position of the rHeB branches.  The histogram displays the [Fe/H] values for
45 EDD dwarfs.  The LSB galaxies are labeled by their names.  All three galaxies
display much lower [Fe/H] values (i.e. bluer rHeB branches) than other nearby dwarfs.
}
\label{metal}
\end{figure}

The current metallicity can also be extracted from the mean position of the rHeB
branch.  As can be seen in Figure \ref{iac}, the rHeB branch moves redward with
increasing [Fe/H].  Calibrating the position of the rHeB branch using synthetic
CMD's, we can assign a current metallicity to each CMD ([Fe/H]$_{rHeB}$).  The
results are shown in Figure \ref{metal} where the histogram displays the deduced
[Fe/H]$_{rHeB}$ values for 45 LV dwarfs with strong rHeB populations.  The three LSB
galaxies in our sample are also marked in Figure \ref{metal} with [Fe/H] values of
$-$1.0, $-$0.6 and $-$0.7 respectfully.  This places all three on the low end of the
distribution, in line with a history of inhibited star formation and, therefore,
a suppressed chemical evolutionary path.  This is also in agreement with the typical
oxygen abundances deduced from LSB emission lines (McGaugh 1994).

\subsection{$U-V$ CMDs}

\begin{figure}[!ht]
\centering
\includegraphics[scale=0.9,angle=0]{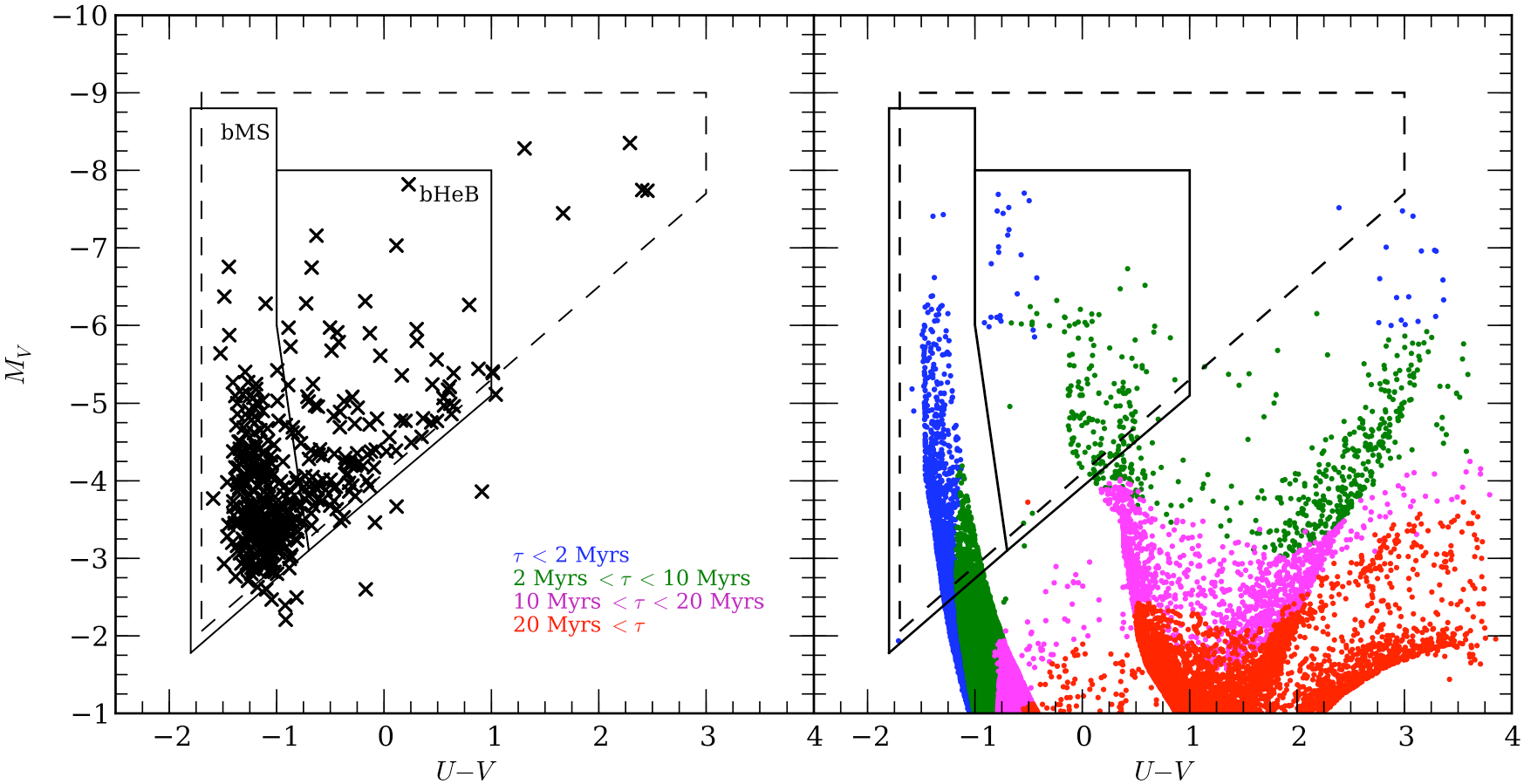}
\caption{\small The $U-V$ CMD for F415-3 compared to an IAC-STAR simulation of
enhanced recent star formation ([Fe/H]=$-0.4$).  The completeness, bMS and bHeB
regions are marked.  The bMS and bHeB branches (measuring
2 Myrs and 10 Myrs stars respectfully) are clearer in the $U-V$ plane than $V-I$ and
the 
ratio of the bMS and bHeB stars will measure recent star formation on timescales of
two to 10 Myrs.
}
\label{uv}
\end{figure}

The bMS region of a $U-V$ CMD is relatively insensitive to metallicity effects as
increasing metallicity primarily lowers the peak luminosity of the brightest O stars,
and the range of blue star absolute luminosity.  Age dominates the position of the
isochrones on the blue side in the $M_V$ versus $U-V$ diagram (see Figure \ref{uv}).
And, unsurprisingly, the bright portion of the blue branch of the CMD can only be
explained by very young ($\tau < 5$ Myrs), metal-poor stars.  A majority of the blue
stars are concentrated along this young isochrone with only 10\% have $U-V$ colors
redder than 0.

Where the bMS and bHeB branches are blurred in the $V-I$ CMD due to photometric
errors, the bHeB branch separates nicely in $U-V$.  Shown in the right panel of
Figure \ref{uv} is an IAC-STAR simulation of a constant star formation, [Fe/H]=$-0.4$
population.  The $U-V$ colors, sampled by our survey, explore the stellar populations
with ages less and 10 Myrs (i.e., very recent), and the ratio of the bMS versus bHeB
regions is a measure of the fraction of 2 Myrs to 10 Myrs stars.  Increasing
metallicity will lower the fraction of bHeB stars (they become fainter and drop below
the completeness line).

Using this $U-V$ CMD diagnostic, we find that the three LSB galaxies in our sample
have bMS and bHeB fractions of 70\% and 25\% on average.  Enhanced recent star
formation predicts fractions of 85\% and 10\% for metal-poor populations.  Constant
star formation models predict 80\% and 20\%.  This would suggest, as indicated in the
previous section, that despite the large fraction of young stars (see Figure
\ref{fractions}) this does not signal a sharp increase in the SFR in the last 50
Myrs.  Rather, this is an indication that the past SFR has been inhibited
such that the blue colors of LSB galaxies is from a suppressed old population, rather
than a recent enhanced cycle of current SF.

\section{Discussion}

Although, due to the distance to our LSB galaxies, their CMD's to not reach to
absolute limiting magnitudes comparable to other LV dwarfs, the LSB CMD's have many
of the same CMD features that LV dwarfs display.  In particular, strong signatures of
recent star formation with numerous OB stars, very low current [Fe/H] values as
deduced by the position of the rHeB population and a measurable deficiency of intermediate
age AGB stars (compared to LV dwarfs).

Our analysis can be divided into three sections; 1) pure observables from the spatial
and color distribution of the LSB CMD's, 2) empirical comparison to CMD's in other
dwarf galaxies and 3) examination of the results from comparison to synthetic CMD
simulations.

\subsection{H$\alpha$ emission and Mean Surface Brightness}

The clearest result from HST stellar photometry of our three LSB galaxies is the
significant one-to-one correspondence between the types and luminosity of the
resolved stars and global features such as local surface brightness, local color and
H$\alpha$ emission.  While this was not unexpected, it is direct confirmation that
the same star formation processes that dominates normal spirals and irregulars are
also found in LSB galaxies (Helmboldt \etal 2009).

For every HII region identified in Schombert, McGaugh \& Maciel (2013) there exists
at least one, often several, stars with $U-V$ colors less than $-$1.  In addition,
the brighter the HII region, the brighter the ionizing stars.  Several groupings are
identified without H$\alpha$ emission (also identified in ground-based imaging as
surface brightness knots) and these regions have a higher fraction of rHeB stars,
i.e.  older than 10 Myrs stars and non-ionizing.  The connection between bright OB
stars and ionized gas confirms that star formation in LSB galaxies proceeds in the
same fashion as normal spirals and irregulars, i.e., collapse of a gas cloud, star
cluster formation, massive star gas ionization followed by gas blowout.   There is no
support of earlier speculation that LSB galaxies form stars without massive stars
(Meurer \etal 2009; Dopcke \etal 2013) nor that H$\alpha$ emission in LSB galaxies is
due to an exotic ionizing population (e.g., blue HB stars or hot white dwarfs).

In addition, the local colors (optical and near-IR) are in direct correspondence with
the colors and luminosity of the local brightest stars.  Regions that are blue in
mean color are also rich in blue stars.  High surface brightness regions are also dominated
by the brightest stars (both blue and red).  Blue regions with low mean surface
brightness have an excess of faint, widely dispersed bMS and bHeB stars suggesting
strong kinematic mixing on short timescales in LSB galaxies.  This is in agreement
with typical gas velocity dispersion estimates of 8 km/sec (Kuzio de Naray \etal
2006) that corresponds to stellar motions of 5 pc per Myr, most than sufficient to
scatter O stars from their original regions of intense star formation.

Lastly, the total luminosity of the resolved population is roughly 10\% of the total
luminosity of the galaxy.  This agrees with the estimates from simulations of
synthetic CMD's, where the ratio between the completeness region and the fainter
stars was between 5 and 20\%, highly variable due to small number statistics of the
brightest stars.  In addition, the stellar counts per pc$^2$ are in excellent
agreement with the local mean surface brightnesses when scaled to the total
luminosity of the galaxy (see Figure \ref{sfb_map}).  This implies there are no
hidden stellar populations in LSB galaxies, the resolved bright stars trace the same
structure as the underlying stellar contribution.  As suspected from the lack of CO
and far-IR detections, LSB have almost no extinction or significant absorption over
the scale sizes of the large star-forming regions (Lynn \etal 2005; Hinz \etal 2007).

The difference between the lowest surface brightness regions and the higher surface
brightness knots is due primarily to the brightest blue stars.  There are numerous B
stars in low surface brightness regions indicating that their ages differ only by a
few tens of Myrs from the bright cluster regions.  In other words, there are no
distinct old regions in LSB galaxies, strong mixing with the star forming regions is
implied or there are simply no obvious regions with stars older than 5 Gyrs,
in conflict with the observed chemical evolution.

\subsection{Comparison to LV dwarfs CMDs}

The $V-I$ CMD has been explored for dozens of LV dwarf galaxies, some as deep as the
turn-off point, but all fainter than the limiting magnitude of our three LSB
galaxies.  The CMD features of LV dwarfs varies widely as their star formation
histories range from very little recent star formation (e.g., IC 3104 and DDO88) to
galaxies with a full range of main sequence, post-main sequence, RGB and post-RGB
features.  In terms of general features, our best CMD in F415-3 contains all the same
CMD features as those in LV dwarfs such as IC 2574 (see Figure \ref{ic2574}).  In
particular, we observe the bMS, bHeB, rHeB and AGB populations.  Our CMD's do not
extend significantly below the tip of the RGB to fully sample the red clump or lower RGB
populations.

Compared to 57 LV dwarf CMD's, we find that the fraction of bMS stars is much higher
in our LSB galaxies than LV dwarfs.  Star forming LV dwarfs have bMS fractions
between 5 and 20\%, whereas LSB galaxies have bMS fractions greater than 30\% (see
Figure \ref{fraction_hist}).  Despite the high fraction of bright blue stars, the
total numbers are in agreement with H$\alpha$ fluxes.  For example, in F415-3,
log $L_{H\alpha}$ is 38.5, which is the equivalent to a cluster slightly larger than
$10^5$ $M_{\sun}$ cluster.  With a normal IMF, this population would have between 800
and 1,000 stars brighter than $M_I = -3$.  In F415-3, there are 850 stars brighter
than this luminosity, which we interpret that there is nothing particularly unusual
on the upper end of the IMF in LSB galaxies.  This is in agreement with the
one-to-one correspondence found between H$\alpha$ emission and the ionizing
population in LV dwarfs (McQuinn \etal 2010), but in contradiction with the
observations of Meurer \etal (2009) who found a deficiency in the upper mass of the
IMF for LSB galaxies (see also Lee \etal 2004).

However, this high bMS fraction must be reconciled with the extremely low SFR rates
for LSB galaxies, typically less than $10^{-3}$ $M_{\sun}$ per yr.  Since the current
SFR is low, the only way to produce a high bMS fraction is to suppress the fraction
of stars in the older populations.  In other words, the stellar population in LSB
galaxies appear to be predominately very young with an underpopulated stellar
population older than 2 to 3 Gyrs.

This is confirmed by the fraction of AGB stars in LSB galaxies, a measure of
intermediate age populations.  For LV dwarfs, the fraction of AGB stars ranges from
20 to 30\% for non-star forming dwarfs to 10\% for star forming dwarfs.  The LSB
galaxies have AGB fractions below 10\%, indicating a much lower SFR in the distant
past which, of course, is in agreement with their abnormally low stellar densities.
This is even a more abnormal fraction for the LSB galaxies in our study are
particularly low in [Fe/H], which should strengthen the AGB population fraction.
Thus, the abnormally blue colors of LSB galaxies is due as much to an absence of old
red stars as much as an overabundance of young blue stars.

There is very little evidence of any stellar population older than 5 Gyrs; however,
our data does not sample the RGB where these stars would lie in the CMD.  A unusually
low fraction of older stars is deduced from the lack of their color signatures
in broadband imaging (Pildis, Schombert \& Eder 1997) and the close correspondence
between the resolved stars and the underlying colors.  Some older population must
exist for the metallicity values, while low, imply the existence of some earlier
enriching stars (see below and, for example, old globular clusters are found in large
LSB galaxies, Villegas \etal 2009).

While a high fraction of bMS and rHeB stars implies either recent surge of SFR 
or a highly suppressed SFR in the past, these conclusions uncomfortably imply that we
live in a special epoch with respect to the star formation history of LSB galaxies.
That is, we are seeing their first epochs of increasing star formation from a large
reservoir of gas reserves.  This is possible, as our sample size is small and
selected from a survey of blue PSS-II plates.  However, more likely, either due to
internal or external inhibitors, we are seeing a global history of steady, but very
slowly increasing star formation where surface brightness is an aftereffect of a very
low, past total star formation rate.  Once a galaxy has achieved a certain value of
SF per pc$^2$, the mean surface brightness of the galaxy exceeds a visibility
threshold, the galaxy becomes detectable for our surveys and catalogs.  Confirmation
of this idea would be the detection of numerous pure-HI systems with very little past
star formation and, therefore, extremely low surface brightness (Davies \etal 2004).

\subsection{Comparison to Synthetic CMDs}

To extract the star formation and chemical evolution history from CMD's one must make
statistical comparisons to artificial CMD's generated with known metallicity and SFRs
as a function of population age.  Several examples are shown in Figure \ref{iac}.
The two regions where comparison to synthetic CMD's is most informative is the rHeB
(see \S2.8) and the $U-V$ CMD (as there are very few $U-V$ CMD's in the literature).

\begin{figure}[!ht]
\centering
\includegraphics[scale=0.9,angle=0]{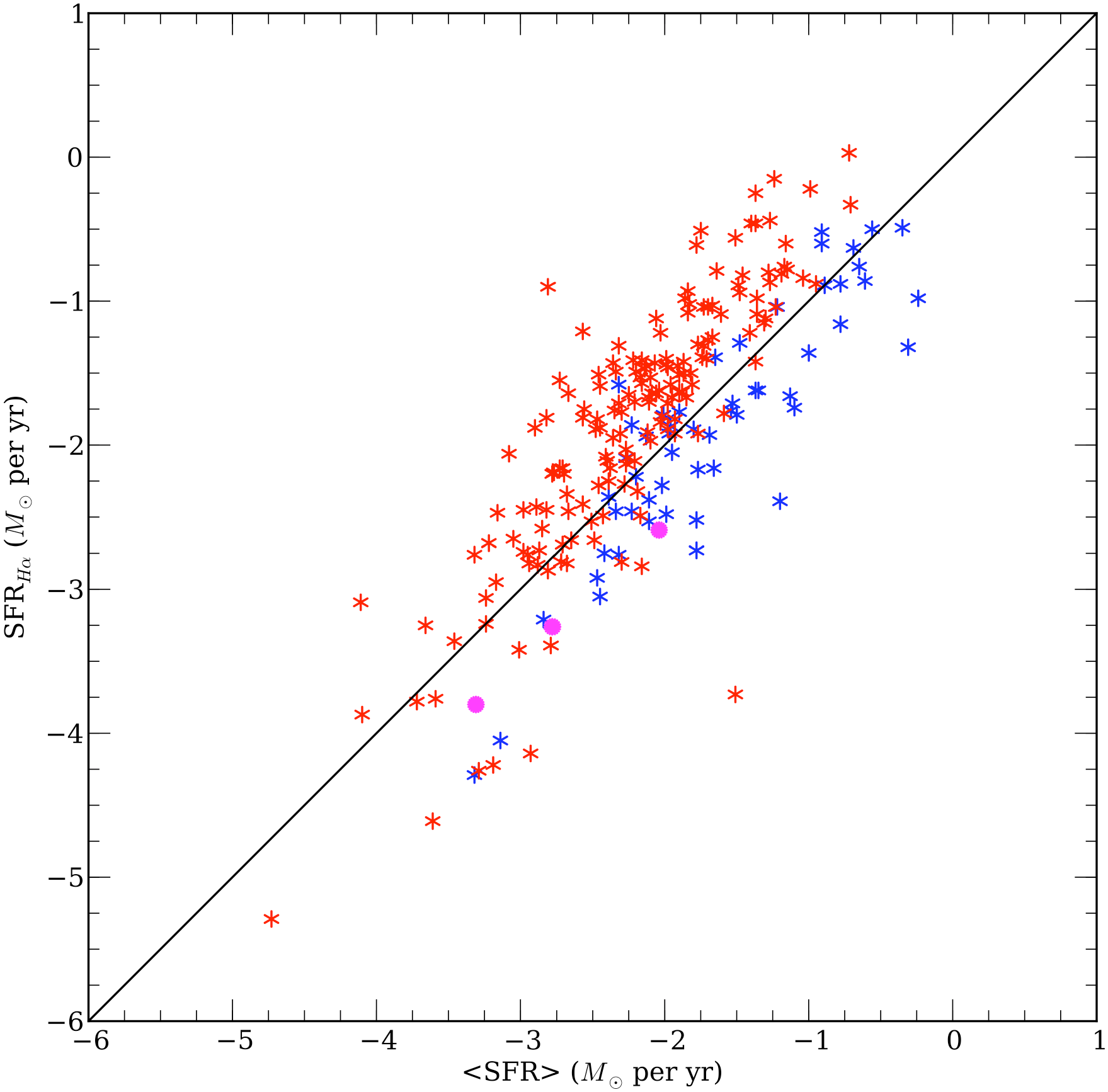}
\caption{\small The mean past rate of star formation, $<$SFR$>$, the stellar mass
divided by 12 Gyrs versus the current SFR from H$\alpha$ luminosities (in $M_{\sun}$
per yr).  The HSB samples of van Zee (2001) and Hunter \& Elmegreen (2004) are shown
as red symbols, LSB galaxies are blue, the three LSB galaxies in this study are
magenta.  HSB irregulars display higher SFRs compare to past rates (explaining their
brighter surface brightnesses) whereas LSB galaxies typically have lower current SFRs
compared to past rate, in conflict with their missing AGB populations.
}
\label{sfr}
\end{figure}

The results we deduce from the rHeB region is that LSBs have very nearly constant SF
for the past 10$^8$ years, slightly stronger than the typical LV dwarf, but well
within the range of recent SFR for LV dwarfs of a range of surface brightnesses.  We
note that although the star formation has been nearly constant, this constant rate is
still at extremely low absolute levels.  Proposing even lower SFRs in the past is in
conflict with the deduced mean $<$SFR$>$ from the stellar mass of HSB and LSB
galaxies.  For comparison, Figure \ref{sfr} displays the current SFR in three samples
of irregular galaxies (our LSB sample; van Zee 2001; Hunter \& Elmegreen 2004) versus
the stellar mass of each galaxies divided by 12 Gyrs (a measure of the mean SFR,
$<$SFR$>$, in a galaxy where galaxy luminosity is converted to stellar mass with a
M/L value.  The M/L values used were deduced by McGaugh \& Schombert (2015) modified
for color following the prescription given in de Blok \& McGaugh (1997) and varied,
at most, from 0.4 to 0.6.  Per unit mass, the LSB galaxies are typically a factor of
ten lower in current SFR than other HSB irregulars.  And their $<$SFR$>$ values are
typically higher than their current SFR values, indicating past rates that are higher
than the current value (the unity line is shown in Figure \ref{sfr} where most HSB
irregulars are above the line and LSB galaxies are below the line).

The global properties of LSB galaxies (compared to HSB irregulars) are difficult to
reconcile with the deficiency of AGB stars in our three LSB galaxies.  While we
lack resolution of truly old stars on the RGB, a deficiency in AGB stars with a
deduced higher past SFRs from Figure \ref{sfr} induces a contrived star formation
history (one with an initial burst sufficient to produce most of the current stellar
mass, then a long quiescent period, to a current epoch of slowly increasing SFR).
While this may very well be the case, and is consistent with low stellar density
distribution, the mechanism for this type of star formation history, given the
stochastic appearance of current star formation, would require some external process
to moderate the quiescent phases (McQuinn \etal 2015).

Second, the position of the rHeB sequence in the $V-I$ CMD is very sensitive to
metallicity of the younger stars.  Calibrating the position to [Fe/H] (using a
standard enrichment scenario), we find the current [Fe/H] values for our three LSB
galaxies are $-$1.0, $-$0.6 and $-$0.7 respectfully.   This is on the low side for
[Fe/H] of LV dwarfs by the same method (their mean value is $-$0.4).  Given the
assumption of lower past SFR's in LSB galaxies compared to LV dwarfs, this is an
unsurprising result and reflects the abnormally low current [Fe/H] values, i.e., the
chemical history of LSB galaxies is strongly suppressed.

Lastly, the $U-V$ CMD allows for a comparison of the bMS and bHeB populations which
are blurred in the $V-I$ CMD.  The ratio of these populations, compared to synthetic
CMD's, confirms the result from the rHeB population, i.e., that the current SFR in
LSB has been roughly constant for the last few 100 Myrs.  Constant star formation is
not a new conclusion for gas-rich galaxies (West \etal 2009; Hunter \etal 2011) and
the bluest gas-rich galaxies require rising SFR to explain their global colors.
Schombert \& McGaugh (2014a) found that recent weak bursts on timescales of 500 Myrs
would satisfy the colors of LSB galaxies, rather than a uniformly rising SFR (see
also Boissier \etal 2003).  It may be a coincidence that the three LSB galaxies in
this sample display the rHeB population indicated the onset of a recent burst (thus,
making them more visible in a blue oriented visual survey).

\section{Summary}

The results from the CMD's of Local Volume dwarf galaxies has historically been a
shockingly revelation on the stochastic and random nature to the star formation
history in dwarf galaxies.  The uniform nature of their global colors and H$\alpha$
luminosities with mass (Hunter \& Elmegreen 2004) is replaced with a highly variable
history of brief, weak bursts.  While our study lacks the luminosity depth and high
number of resolved stellar sources to match the detail of most LV dwarfs CMD's, the
similarity between the CMD's for LSB galaxies and LV dwarfs indicates that they have
analogous recent star formation histories.

The primary difference between LV dwarfs and our LSB galaxies is the underpopulated
older population ($\tau >$ 3 Gyrs), implied by the overabundance of young stars and,
yet, a low current SFR.  Most studies of LSB galaxies in the past have tested and
dismissed various explanations for their LSB nature based on stellar population
variations (extinction by dust, unusual IMF's, exotic stellar populations).  Whereas,
this study indicates that that LSB are low in surface brightness simply because they
have lower stellar densities due to a widely dispersed stellar population.  In other
words, they have fewer stars per pc$^2$ than their HSB cousins, and that this
underpopulation occurs with both recent and older stars.  Their current and past star
formation rates are typically a factor of ten less then their HSB cousins, which
clearly reflects into their different mean surface brightness.  However, a kinematic
mechanism is required to disperse the younger stellar populations to maintain the
uniform color mixing from high to low surface brightness regions within the LSB
galaxies themselves.

This resurrects then idea that LSB galaxies are "young" (Vorobyov \etal 2009; Gao
\etal 2010), not necessarily young in their formation epoch, for their mean
metallicities indicate some small amount of chemical evolution over the
that last 10 Gyrs.  Rather they are "young" in the sense that a majority of their
stars formed after 5 Gyrs (McGaugh \& Bothun 1994; Jimenez \etal 1998; Schombert,
McGaugh \& Eder 2001), and their chemical history is the weakest of any galaxy type.
We also note that the analysis of rHeB population in \S2.8 opens a powerful technique
to study galaxies outside the Local Volume out to 20 Mpc or more.  There is a great
deal of information in the resolved stellar populations brighter than $M_I < -4$, the
canonical value for the tip of the RGB.

Ultimately, the conclusions from H$\alpha$ and CMD studies is that star formation is
suppressed in LSB galaxies.  However, the dilemma exists on why should the SFR be so
low in galaxies so rich in the neutral gas that is the fuel for star formation.  The
evidence points to the star formation efficiency as the difference between HSB and
LSB galaxy types.  While star formation has been directly linked to gas density
(Kennicutt 1989), numerous secondary factors vary with surface brightness.  For
example, it has been shown that star formation efficiency decreases with surface
brightness (Leroy \etal 2008) and driven in part by lower metallicities in the gas
clouds (Shi \etal 2014).  This results in a fluctuating (bursts) and a spatially
irregular star formation history (McQuinn \etal 2015) such as seen in most LSB
galaxies.

\acknowledgements Software for this project was developed under NASA's AIRS and ADP
Programs. Based on observations made with the NASA/ESA Hubble Space Telescope, which
is operated by the Association of Universities for Research in Astronomy, Inc., under
NASA contract NAS 5-26555. These observations are associated with program 12859.
This work has made use of the IAC-STAR Synthetic CMD computation code. IAC-STAR is
suported and maintained by the computer division of the Instituto de Astrofsica de
Canarias.

\begin{deluxetable}{lccccccccc}
\tablecolumns{10}
\small
\tablewidth{0pt}
\tablecaption{Optical and HI Properties}

\tablehead{
\colhead{Object} &
\colhead{Distance} &
\colhead{$M_V$} &
\colhead{$B-V$} &
\colhead{$\mu_o$} &
\colhead{$\alpha$} &
\colhead{log $L_{H\alpha}$} &
\colhead{log $M_*$} &
\colhead{log $M_{HI}$} &
\colhead{$f_g$} \\

\colhead{} &
\colhead{(Mpc)} &
\colhead{} &
\colhead{} &
\colhead{(V)} &
\colhead{(kpc)} &
\colhead{(ergs s$^{-1}$)} &
\colhead{($M_{\sun}$)} &
\colhead{($M_{\sun}$)} &
\colhead{} \\

}

\startdata

F415-3 (UGC 2017) & 10.4 & -15.2 & 0.52 & 22.8 & 0.7 & 38.5 & 8.04 & 8.65 & 0.80 \\
F608-1 (UGC 159)  &  9.0 & -13.5 & 0.50 & 23.7 & 0.4 & 37.8 & 7.30 & 7.73 & 0.73 \\
F750-V1           &  8.0 & -12.7 & 0.32 & 22.7 & 0.2 & 37.3 & 6.77 & 7.14 & 0.70 \\

\enddata
\end{deluxetable}


\end{document}